\documentclass[prl,letterpaper,aps,10pt,superscriptaddress,twocolumn,floatfix,showpacs]{revtex4-1}
\usepackage{graphicx}
\usepackage{amsmath}
\usepackage{amsfonts}
\usepackage{amssymb}
\usepackage{epsfig}
\usepackage[pdftex]{color}
\usepackage{epstopdf}
\usepackage{amsmath,graphicx,amssymb,braket,subfigure,upgreek}
\usepackage[colorlinks=true, linkcolor=blue, citecolor=blue, urlcolor=blue, breaklinks=true]{hyperref}
\usepackage{bigints}
\usepackage{dcolumn}
\usepackage{bm}
\usepackage{mathtools}
\usepackage{mathrsfs}
\usepackage{frcursive}
\usepackage{dsfont}
\usepackage{bbm}

\begin{document}
\date{\today}
\title{Cavity-enhanced transport of charge}

\author{David Hagenm\"{u}ller}
\affiliation{IPCMS (UMR 7504) and ISIS (UMR 7006), University of Strasbourg and CNRS, 67000 Strasbourg, France}
\author{Johannes Schachenmayer}
\affiliation{IPCMS (UMR 7504) and ISIS (UMR 7006), University of Strasbourg and CNRS, 67000 Strasbourg, France}
\author{Stefan Sch\"utz}
\affiliation{IPCMS (UMR 7504) and ISIS (UMR 7006), University of Strasbourg and CNRS, 67000 Strasbourg, France}
\author{Claudiu Genes}
\affiliation{IPCMS (UMR 7504) and ISIS (UMR 7006), University of Strasbourg and CNRS, 67000 Strasbourg, France}
\affiliation{Max Planck Institute for the Science of Light, Staudtstra{\ss}e 2, D-91058 Erlangen, Germany}
\author{Guido Pupillo}
\affiliation{IPCMS (UMR 7504) and ISIS (UMR 7006), University of Strasbourg and CNRS, 67000 Strasbourg, France}

\date{\today}
\begin{abstract}
We theoretically investigate charge transport through electronic bands of a mesoscopic one-dimensional system, where inter-band transitions are coupled to a confined cavity mode, initially prepared close to its vacuum. This coupling leads to light-matter hybridization where the dressed fermionic bands interact via absorption and emission of dressed cavity-photons. Using a self-consistent non-equilibrium Green's function method, we compute electronic transmissions and cavity photon spectra and demonstrate how light-matter coupling can lead to an enhancement of charge conductivity in the steady-state. We find that depending on cavity loss rate, electronic bandwidth, and coupling strength, the dynamics involves either an individual or a collective response of Bloch states, and explain how this affects the current enhancement. We show that the charge conductivity enhancement can reach orders of magnitudes under experimentally relevant conditions.
\end{abstract}


\maketitle

% PLACEMENT OF PAPER

The study of strong light-matter interactions~\cite{weisbuch,lidzey,raimond,reithmaier} is playing an increasingly crucial role in understanding as well as engineering new states of matter with relevance to the fields of quantum optics~\cite{yoshie,savasta,fushman,tsintzos,lagoudakis,claudon,kenacohen,cristofolini,albert,schneider,yalla,goban,javadi,sipahigil}, solid state physics~\cite{tredicucci,kasprzak,hennessy,amo,deng,carusotto,menard,imam1,imam2,imam3,liu_semiconductor_2015, frey_dipole_2012,viennot_out--equilibrium_2014}, as well as quantum chemistry~\cite{ebbesen,jino,herrera,baumberg,galego} and material science~\cite{badolato,tischler,akimov,sapienza,aberra,vasa,tudela,tanese,ballarini,bhattacharya,plumhof,shi,xiaoze,tumkur,keeling,lacount,grant}. An emerging topic of interest is the modification of material properties using either external electro-magnetic radiation~\cite{mitrano,yamamoto,mazza} or spatially confined modes such as in cavity quantum electrodynamics~\cite{liberato,laussy,morina,feist2015extraordinary,schachenmayer2015cavity,kavokin,gudmundsson}. Recent experiments with organic semiconductors have demonstrated a dramatic enhancement of charge conductivity when molecules interact strongly with a surface plasmon mode~\cite{orgiu2015conductivity}. In principle, this can open up exciting new opportunities both for basic science and applications of organic electronics~\cite{salleo}. The microscopic mechanisms leading to charge conductivity enhancement, however, remain today largely unexplained. In this work, we  propose a  proof-of-principle model that sheds light on the physical mechanisms behind current enhancement due to the interaction with a confined bosonic mode. We show that our model can lead to a dramatic current enhancement by orders of magnitude for certain conditions that can be relevant to typical experiments across fields.

% SUMMARY OF RESULTS WITH EASY INTRO TO MODEL

\begin{figure}[t]
\centerline{\includegraphics[width=0.85\columnwidth]{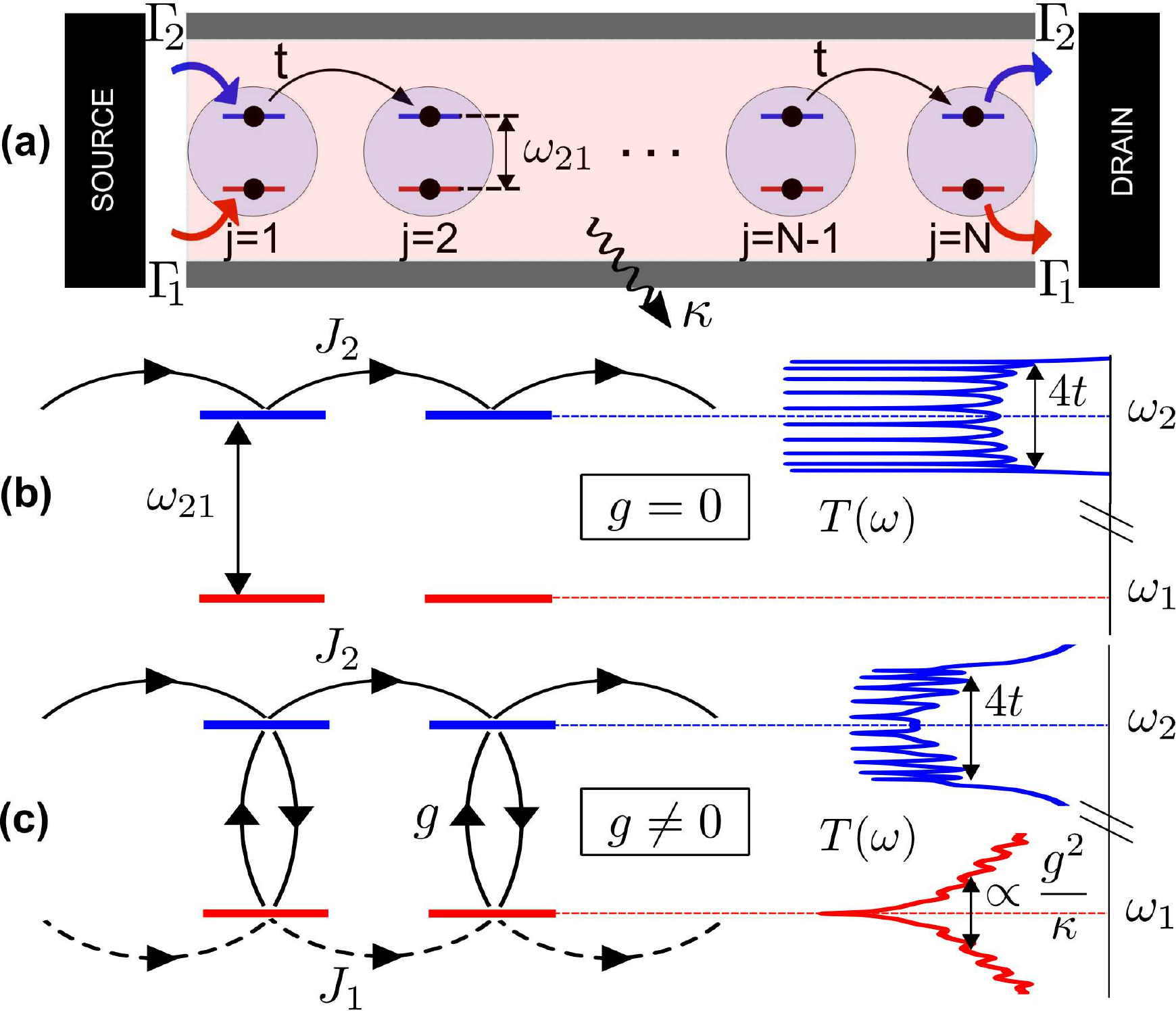}}
\caption{\textbf{(a)} Model for 1D charge transport in the presence of a cavity. \textbf{(b)} In the absence of light-matter coupling, the hopping $t$ allows for  transmission in the upper band only. \textbf{(c)} In the presence of light-matter coupling (coupling strength $g$, photon loss rate $\kappa$), current can effectively flow through the two dressed bands (the currents $J_1$ and $J_2$ are defined in the text) providing a new contribution of width $\propto g^{2}/\kappa$ in the transmission spectrum $T(\omega)$.}
\label{fig1}
\end{figure}

The setup we consider [see Fig.~\hyperref[fig1]{\ref*{fig1}(\textbf{a})}] consists of a mesoscopic chain of $N$ sites with two orbitals of energy $\omega_1$ and $\omega_2$ ($\hbar=1$) in a 1D geometry, forming two bands in a tight-binding picture. The edges of the chain are connected to a source and a drain with a bias voltage across, respectively inserting and removing (spin-less) electrons in the two orbitals at rates $\Gamma_{1}$ and $\Gamma_{2}$. The on-site interband transitions with energy $\omega_{21}=\omega_2-\omega_1$ are resonantly coupled to a cavity mode with coupling strength $g$ and loss rate $\kappa$. Initially, we only allow electrons in the upper band to hop with a rate $t_2 \equiv t$, while no hopping is assumed in the lower band ($t_1= 0$). Furthermore, we start with a situation of a large bias voltage, such that the Fermi level of the source (the drain) is higher (lower) than any other energy scale in the system, allowing for injection/extraction in both bands at a rate $\Gamma_1 = \Gamma_2 \equiv \Gamma$.

 In the case $g=0$ [Fig.~\hyperref[fig1]{\ref*{fig1}(\textbf{b})}] electronic transmission can only arise due to the bare upper-band Bloch states. We find that for $g\neq 0$ [Fig.~\hyperref[fig1]{\ref*{fig1}(\textbf{c})}], the states of the two bands and the cavity mode hybridize to new states with enhanced transmission properties. Effectively, this results from a restoration of tunneling through the previously blocked lower band. More precisely, we show that the current enhancement is determined by the photon spectral weight present within the electronic bandwidth, which enables ``individual'' hybridization between the lower and the upper band Bloch states. In contrast to the usual Tavis-Cummings (TC) model for spins~\cite{tavis}, where the system properties are determined by the ratio $g/\kappa$ only, here, the nature of the light-matter coupling and that of the current enhancement also crucially depends on the ratios $\kappa/4t$ and $g/4t$ ($4t$ is the electronic bandwidth). We show that when $\kappa,g\ll4t$, the hybridized states retain a well-defined quasi-momemtum, and the current enhancement can be interpreted as an effective hopping mechanism [sketched in Fig.~\hyperref[fig1]{\ref*{fig1}(\textbf{c})}]. However, when the band dressing becomes collective (e.g. for $\kappa\gg4t$, or $\kappa\ll4t$ and $g>4t$), damped oscillations of the charge density between the two bands associated to polariton states play an important role, a process that does not contribute to inter-site charge transport. Ultimately, when the polariton splitting becomes much larger than $4t$, the current enhancement vanishes and one recovers the TC physics. Finally, we show that the characteristics mentioned above can be identified in the cavity photon spectrum and could be directly accessed in absorption spectroscopy experiments.

The steady-state current $J$ can be computed through the electronic transmission spectrum $T(\omega)$ as
\begin{equation}
J =\frac{e \Gamma}{2} \int \!\! \frac{d\omega}{2\pi} T(\omega),
\label{final_current_sus_long}
\end{equation}
with $\omega$ the frequency and $e$ the electron charge. The dressing of the electronic bands and the cavity mode requires a self-consistent solution of the problem, which we obtain using a non-equilibrium Green's function method~\cite{caroli1,caroli2,damato,datta,kim,roy,haug,pourfath,doornenbal,gruss,SM}. For $g\neq 0$, we show that $T(\omega)$ acquires a new additional transmission channel [sketched in Fig.~\hyperref[fig1]{\ref*{fig1}(\textbf{c})}], which is responsible for the current enhancement.

We consider the Hamiltonian $H_{S}=H_{\rm el} + H_{\rm int}+H_{\rm cav}$, where
\begin{align}
&H_{\rm el} = \sum_{j=1}^{N} \sum_{\alpha=1}^{2} \omega_{\alpha} c^{\dagger}_{\alpha,j} c_{\alpha,j} - t \sum_{j=1}^{N-1} \left(c^{\dagger}_{2,j+1} c_{2,j} + \rm{h.c.} \right) \label{H_kinetic} \\
&H_{\rm int} = g \sum_{j=1}^{N} \left(c^{\dagger}_{2,j} c_{1,j} a + c^{\dagger}_{1,j} c_{2,j} a^{\dagger} \right),
\label{H_intera}
\end{align}
and $H_{\rm cav}=\omega_{21} a^\dagger a$. Here, $a$ is the bosonic annihilation operator for the cavity mode, while the fermionic operator $c_{\alpha,j}$ annihilates an electron in the orbital $\alpha=1,2$ on site $j$. The term $H_{\rm el}$~\hyperref[{H_kinetic}]{(\ref*{H_kinetic})} is diagonalized in $k$-space as $H_{\rm el}=\sum_{\alpha,k} \omega_{\alpha,k} \tilde{c}^{\dagger}_{\alpha,k} \tilde{c}_{\alpha,k}$, with $\omega_{2,k} = \omega_{2} - 2 t \cos \left( \pi k/(N+1) \right)$ and $\omega_{1,k} = \omega_{1}$. The Hamiltonian $H_{S}$ can be thus partitioned into a diagonal part $H_{0}=H_{\rm el}+H_{\rm cav}$ with known eigenstates, and the light-matter interaction~\hyperref[H_intera]{(\ref*{H_intera})} which is treated perturbatively. In the following, all energies are in units of $\omega_{21}$, which is set to $1$. In the high-bias regime, the transmission function entering Eq.~\hyperref[final_current_sus_long]{(\ref*{final_current_sus_long})} is derived as~(see Supplemental Material~\cite{SM})
\begin{align}
T(\omega) &= \textbf{Tr} \left[\underline{\sigma}^{1} \circ \underline{A}_{\alpha} (\omega) + \left( \underline{\sigma}^{N} -\underline{\sigma}^{1}\right) \circ \Im  \underline{G}^{<}_{\alpha} (\omega) \right].
\label{trans_long_coco}
\end{align}
Here, $\textbf{Tr} \equiv \sum_{\alpha,k,k'}$, underlined quantities denote $N\times N$ matrices, $\circ$ is the element-wise Hadamard product, $\Im$ stands for imaginary part, and $\underline{\sigma}^{j}$ is a matrix of Fourier coefficients~\cite{SM}. Equations~\hyperref[final_current_sus_long]{(\ref*{final_current_sus_long})} and \hyperref[trans_long_coco]{(\ref*{trans_long_coco})} correspond to a generalization of the Lan\-dau\-er formula~\cite{landauer} to non-equilibrium mesoscopic systems~\cite{haug,SM}. 

The first contribution of Eq.~\hyperref[trans_long_coco]{(\ref*{trans_long_coco})} involves the trace of the electron spectral function $\underline{A}_{\alpha} (\omega)=- 2 \Im \underline{G}^{r}_{\alpha}(\omega)$ in the band $\alpha$. Here, $\underline{G}^{r}_{\alpha}(\omega)$ is the retarded Green function (GF) with matrix elements $G^{r}_{\alpha,k,k'} (\omega) =-i \int_{0}^{+\infty} \!\! d\tau e^{i \omega \tau} \langle \{ \tilde{c}_{\alpha,k} (\tau), \tilde{c}^{\dagger}_{\alpha,k'} (0) \} \rangle$, where $\langle \cdots \rangle$ corresponds to the expectation value in the steady-state. Physically, the quantity $\sum_{k,k'} {A}_{\alpha,k,k'} (\omega)$  corresponds to the normalized electron density of states (DOS) in the band $\alpha$. Similarly to electrons, a cavity photon DOS $A_c(\omega)$  can be introduced \cite{SM}, which can be directly accessed experimentally by measuring the cavity absorption spectrum.

As implied by the spectral function normalization $\int \! d\omega A_{\alpha,k,k'}(\omega) =2\pi \delta_{k,k'}$, the effect of light-matter interactions on the steady-state current is entirely determined by the second term in Eq.~\hyperref[trans_long_coco]{(\ref*{trans_long_coco})}. This contribution is proportional to the trace of the ``lesser'' electron GF, with matrix elements $G^{<}_{\alpha,k,k'} (\omega) =i \int_{-\infty}^{+\infty} \!\! d\tau e^{i \omega \tau} \langle \tilde{c}^{\dagger}_{\alpha,k'} (0) \tilde{c}_{\alpha,k} (\tau) \rangle$. From there, one can compute expectation values $2\pi n_{\alpha,k,k'}=\int \! d\omega \Im G^{<}_{\alpha,k,k'} (\omega)$ as well as real space populations $n_{\alpha,j} = \langle {c}^{\dagger}_{\alpha,j} {c}_{\alpha,j} \rangle$ in the steady-state~\cite{pourfath}. The GFs $\underline{G}^{r}_{\alpha} (\omega)$ and $\underline{G}^{<}_{\alpha} (\omega)$ can be related to electron and photon self-energies through Dyson and Keldysh equations. The latter can be solved perturbatively in the framework of the self-consistent Born approximation, which leads to a closed set of integro-differential equations. In~\cite{SM}, we show in detail how these equations can be solved by iterations. 

In the absence of light-matter coupling ($g=0$), the steady-state current flowing through the upper band is entirely driven by the ratio $t/\Gamma$ \cite{SM}: 
\begin{align}
J = \frac{e\Gamma/2}{1+(\Gamma/2t)^2}.
\label{eq:stb}
\end{align}
When $t \ll \Gamma$, the current vanishes as $J \sim 2et^2/\Gamma$, while it reaches its maximum of $e\Gamma/2$ when $t \gg \Gamma$. In the latter regime, $T(\omega)$ consists of $N$ well-resolved peaks of width $\propto \Gamma$ associated with the different Bloch states~\cite{caroli1,caroli2,kim,dhar,doornenbal} [see Fig.~\hyperref[fig1]{\ref*{fig1}(\textbf{b})}]. In the following, we focus on $t \gg \Gamma$ and $g\neq 0$.

\begin{figure}[t]
\centerline{\includegraphics[width=0.85\columnwidth]{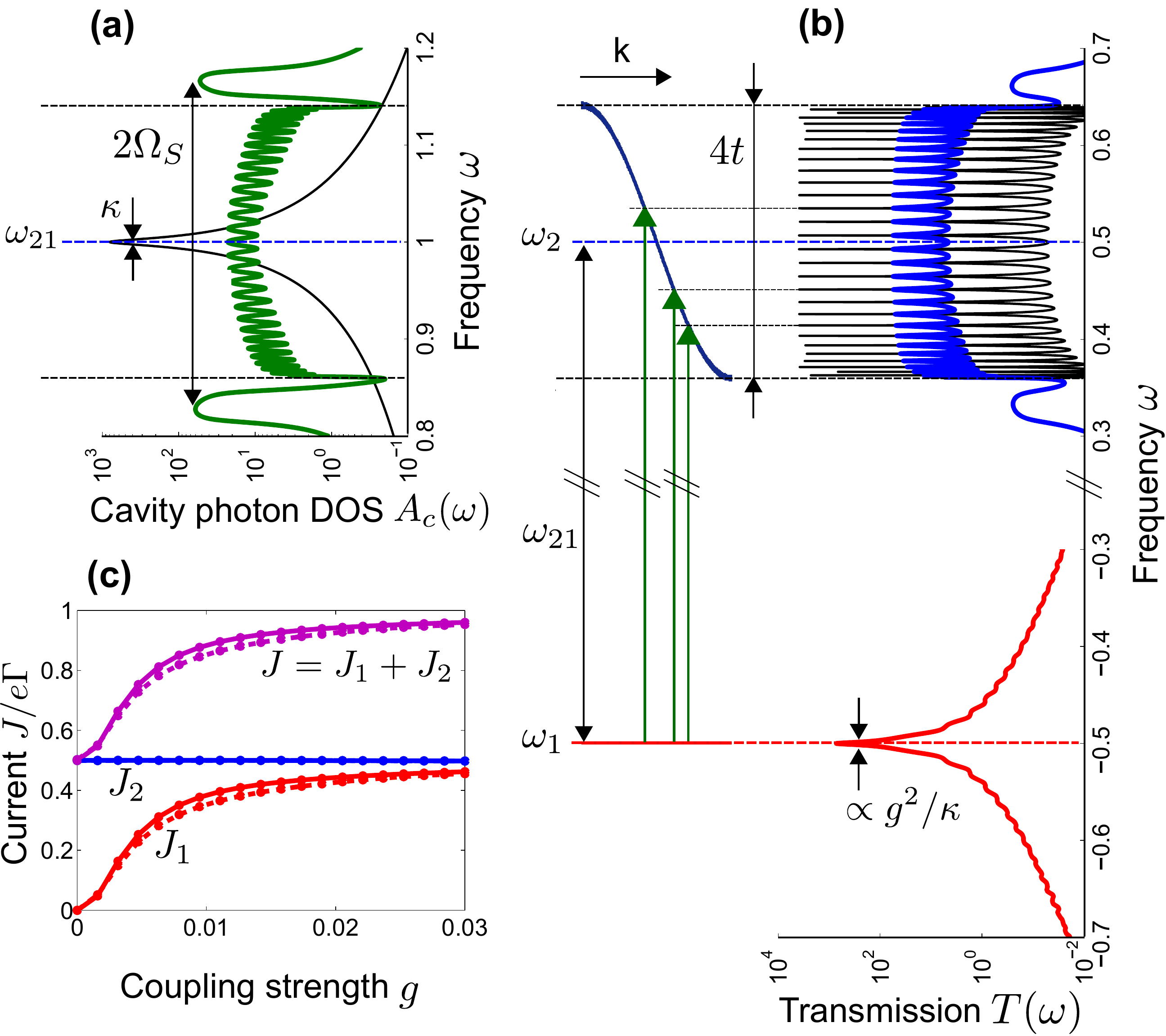}}
\caption{``Individual dressing regime'', with $\Gamma=2.5 \times 10^{-4}$, $t=0.07$, and $\kappa=5 \times 10^{-3}$. \textbf{(a)} (Log-scale) Cavity photon density of state $A_{c}(\omega)$ for $N=30$. The thin black line corresponds to the bare cavity mode ($g=0$), and the green line to the dressed one ($g=0.03$). \textbf{(b)} (Log-scale) Transmission function $T(\omega)$ for $N=30$. The thin black line corresponds to $g=0$, while the red and blue lines correspond respectively to the vicinity of the lower ($\sim \omega_{1}=-0.5$) and the upper band ($\sim \omega_{2}=0.5$) for $g=0.03$. The Bloch states dispersion is plotted on the left side of panel \textbf{(b)}, and a few interband transitions are depicted as vertical green arrows. \textbf{(c)} Currents  versus coupling strength $g$ for $N=10$ (solid) and $N=30$ (dashed). Red, blue, and magenta lines represent the partial currents $J_{1}$, $J_{2}$, and the total current $J$ (see text).}
\label{fig3}
\end{figure}

In the presence of light-matter coupling, we find that the electron DOS is redistributed among the two bands, modifying the transmission spectrum. In particular, $T(\omega)$ acquires a peak of width $\propto g^{2}/\kappa$, centered around the bare lower band energy $\omega_1$. As shown in \cite{SM}, this scaling can be explained  by an analytical calculation up to second order in the perturbation~\hyperref[H_intera]{(\ref*{H_intera})}. Note that our method is exact in the perturbative regime $g^{2}/\kappa \ll \Gamma$, and $\kappa,g \ll \omega_{21}$. Nevertheless, we find qualitatively correct results even for $g^{2}/\kappa >\Gamma$ (as seen by comparisons with master equation simulations \cite{SM}). Using our method we now identify the microscopic mechanisms giving rise to the modified charge transmission properties in two distinct regimes.

We denote the case where $\kappa,g \ll 4t$ as ``individual dressing regime''. This situation is depicted in Fig.~\hyperref[fig3]{\ref*{fig3}} for an example with $N=30$, $\Gamma=2.5 \times 10^{-4}$, $t=0.07$, and $\kappa=5 \times 10^{-3}$. Fig.~\hyperref[fig3]{\ref*{fig3}(\textbf{a})} displays the steady-state cavity photon DOS (for $g=0.03$), which we find to be a key quantity to explore the interplay between charge transport and strong coupling physics. In this regime, the narrow bare cavity mode (thin black line) is resonant with only a few inter-band transitions (with frequencies $\omega_{2,k}-\omega_{1}$) between the lower and the upper band Bloch states. We find that when $g$ is larger than the separation between adjacent Bloch states, two polariton peaks separated by a splitting $2\Omega_S$ appear outside the electronic bandwidth, while inside the bandwidth, we clearly resolve $N-1$ peaks associated to inter-band transitions~\footnote{Some of the qualitative features of the cavity photon DOS follow from a simple generalization of the TC Hamiltonian~\cite{tavis}: $H_{\rm TC}=\omega_{0}a^{\dagger} a+ \sum_{k=1}^{N} (\omega_{2,k}-\omega_{1}) \sigma^{z}_{k}/2 + g \sum_{k} (\sigma^{+}_{k} a+\sigma_{k}^{-} a^{\dagger})$, where $\sigma^{z}_{k},\sigma^{\pm}_{k}$ are the Pauli operators associated with the $N$ inter-band transitions.}.

The transmission spectrum $T(\omega)$ is shown in Fig.~\hyperref[fig3]{\ref*{fig3}(\textbf{b})} for $g=0$ (black line) and $g=0.03$ (blue and red lines). The key feature is the appearance of the large peak of width $\propto g^2/\kappa$ centered at the bare flat band energy $\omega_1$ for $g\neq 0$. This peak originates from inter-band electronic transitions concurrently with the absorption/emission of cavity photons with energy $\omega\approx \omega_{21}$. This new transmission corresponds to the opening of a transport channel with effective hopping rate $\propto g^{2}/\kappa$, responsible for the observed current enhancement. Note that the peaked structure of the upper-band Bloch states still remains visible in $T(\omega)$ for $g\neq 0$. This indicates that a well-defined quasi-momentum can still be associated to the dressed Bloch states, which supports the coherent effective hopping picture. The two polariton peaks from the cavity photon DOS give rise to only marginal contributions outside the bandwidth $4t$ (note the log-scale).

Figure~\hyperref[fig3]{\ref*{fig3}(\textbf{c})} shows the effective currents $J_1$ and $J_2$ as a function of $g$, that are obtained by integrating $T(\omega)$ in the vicinity of $\omega_1$ and $\omega_2$, respectively. The effective lower band current $J_1$ results from the new channel appearing around $\omega_1$ and strongly increases in the considered range of $g$. Crucially, in this individual dressing regime, the current $J_2$ is barely affected by the coupling, and we find that the currents are nearly independent of the chain length $N$. The overall current $J=J_1+J_2$ \footnote{Note that the overall current can be also obtained using a master equation approach as we detail in \cite{SM} We find that the results are consistent with the non-equilibrium Green's function method.} reaches its maximum $e\Gamma$ asymptotically for large $g$.

\begin{figure}[t]
\centerline{\includegraphics[width=0.85\columnwidth]{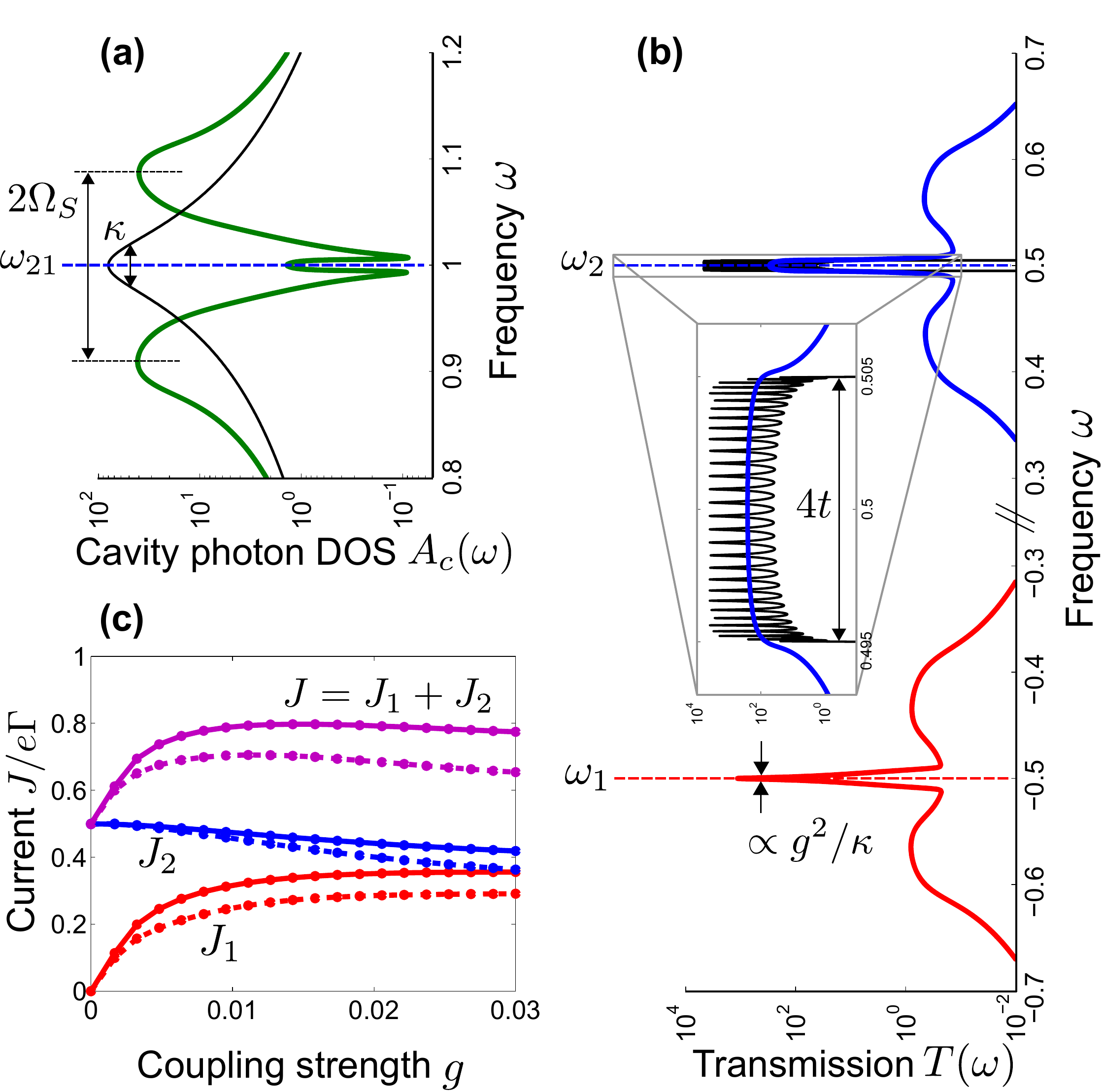}}
\caption{``Collective dressing regime''. Same quantities as in Fig.~\hyperref[fig3]{\ref*{fig3}} for $\Gamma=2.5 \times 10^{-4}$, $t=2.5 \times 10^{-3}$, and $\kappa=0.05$. $N$ and $g$ are identical to Fig.~\hyperref[fig3]{\ref*{fig3}}. \textbf{(a)} (Log-scale) Cavity photon density of state $A_{c}(\omega)$. \textbf{(b)} (Log-scale) Transmission function $T(\omega)$. The central region is shown in the inset. \textbf{(c)} Partial and total currents versus coupling strength $g$.}
\label{fig2}
\end{figure}

%\pagebreak

The ``collective dressing regime'' is typically achieved when $\kappa \gg 4t$. This is depicted in Fig.~\hyperref[fig2]{\ref*{fig2}} for an example with $N=30$, $\Gamma=2.5 \times 10^{-4}$, $t=2.5 \times 10^{-3}$, and $\kappa=0.05$. Figure~\hyperref[fig2]{\ref*{fig2}(\textbf{a})} displays the cavity DOS. Here, for $g=0$, the broad bare cavity mode of width $\kappa$ (thin black line) is resonant to all inter-band transitions. The photon DOS in the coupled case with $g=0.03$ is shown as a thick green line. The small peak centered at $\omega_{21}$ now consists of $N-1$ overlapping peaks (not resolved) with small photon weight. In this situation, we again observe two polariton peaks outside the electronic bandwidth, but in contrast to Fig.~\hyperref[fig3]{\ref*{fig3}(\textbf{a})}, they concentrate most of the photon spectral weight, which reduces the individual band dressing and the current enhancement. Here, the dynamics is dominated by collective oscillations of the charge density at a frequency $\Omega_{S}$, as discussed below.

The transmission spectrum $T(\omega)$ in the steady-state is shown in Fig.~\hyperref[fig2]{\ref*{fig2}(\textbf{b})} for $g=0$ (thin black line, inset) and $g=0.03$ (blue and red lines). Again, we observe large peaks around $\omega_1$ (red) and $\omega_2$ (blue). In contrast to Fig.~\hyperref[fig3]{\ref*{fig3}}, the peaked structure associated to the individual band dressing disappears, indicating a collective response of the Bloch states that can not be associated with a well-defined quasi-momentum (see inset). 

In Fig.~\hyperref[fig2]{\ref*{fig2}(\textbf{c})} it becomes evident that this collective behavior leads to an upper band current $J_2$ decreasing with $g$, and resulting in a maximum in the overall current. This feature can be intuitively understood, as damped collective oscillations remove populations from the upper band. Furthermore, we find that $J$ decreases significantly when increasing the chain length $N$, which is another indication of the presence of collective effects (i.e.~the relevant coupling parameter is $\Omega_S$ and not $g$ \cite{SM}). Ultimately, when $\Omega_S \gg \kappa \gg 4t$, the photon DOS resembles the standard TC spectrum~\cite{tavis} with two well-separated polariton peaks but no photon spectral weight at the inter-band frequencies, thereby preventing the individual band dressing and the current enhancement.

Our open system exhibits a larger Hilbert space ($4^N$) compared to the usual TC model for spins ($2^N$). The latter can be recovered only by constraining the electron number to one for each two-level system, providing a splitting $2 g\sqrt{N}$ between the two polaritons~\cite{tavis}. In our two-band model, a collective vacuum Rabi splitting can be defined as $\Omega_{n}=g\sqrt{N_1 - N_2}$ ($N_\alpha = \sum_{j} n_{\alpha,j}$)~\cite{todorov}, which is obtained from the steady-state population imbalance between the two bands. Importantly, since sites with both orbitals occupied (or empty) are not effectively coupled to light, we always find $\Omega_n < g\sqrt{N}$, which implies that the TC model cannot be used to determine, e.g., the number of molecules based on measured $\Omega_S$ in experiments.  In general, $\Omega_{n}\approx \Omega_S$ is an indication of the presence of collective effects associated to joint charge oscillations between the two bands. This is the case in Fig.~\hyperref[fig2]{\ref*{fig2}}, however, this dynamics can also be recovered in the regime where $\kappa \ll 4t$ and $g > 4t$, i.e. when $g$ is large enough to couple all the Bloch states together (see \cite{SM}).

\begin{figure}[t]
\centerline{\includegraphics[width=1\columnwidth]{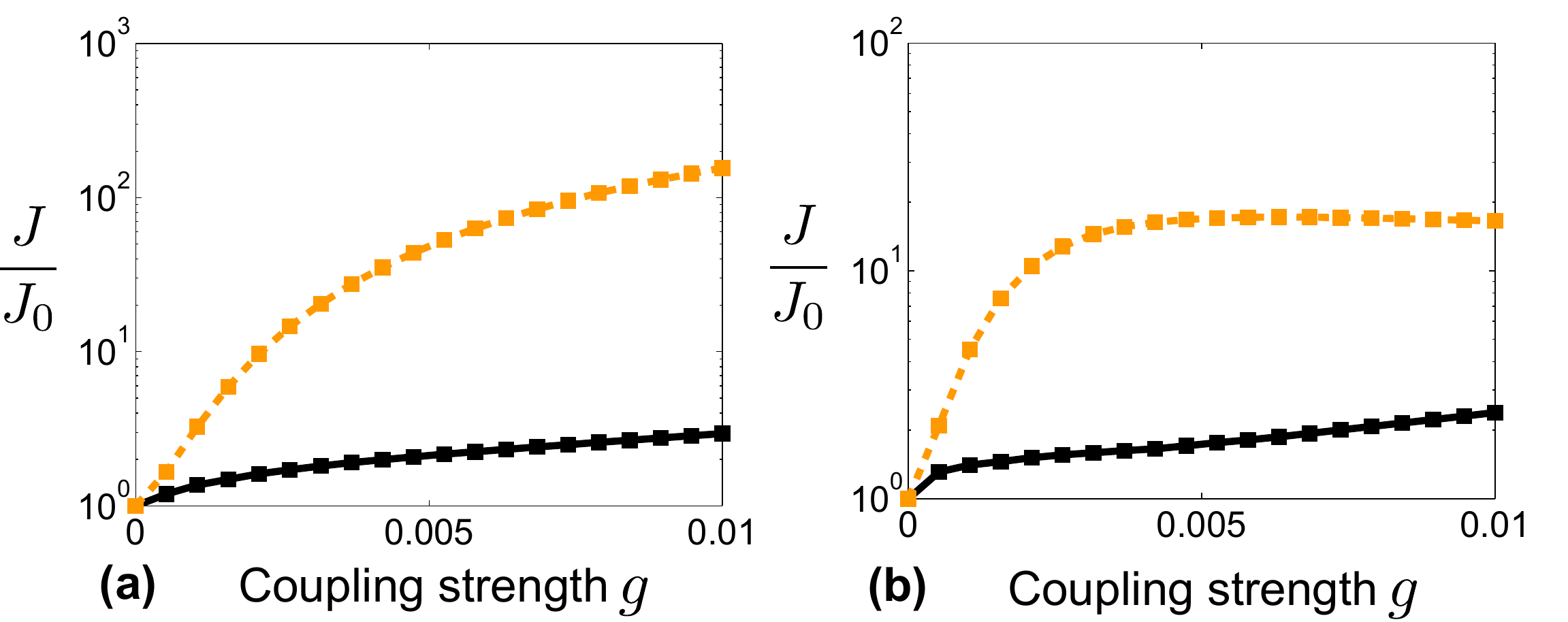}}
\caption{Current enhancement $J/J_{0}$ ($J_{0}$ is the current for $g=0$) versus $g$. \textbf{(a)} ``Individual dressing regime'' with $t=0.07$, $\Gamma_{1}=0.01$, and $\kappa=5 \times 10^{-3}$. \textbf{(b)} ``Collective dressing regime'' with $t=2.5 \times 10^{-3}$, $\Gamma_{1}=10^{-3}$, and $\kappa=0.05$. The full black and dotted orange lines correspond to $N_{\rm ph}=0,0.5$, respectively. Other parameters are $N=30$, $\Gamma_{2}=10^{-5}$, and $t_{1}=5 \times 10^{-5}$.}
\label{fig:phot}
\end{figure}

% realistic setup
Having identified the current enhancement mechanisms, we now consider a scenario more reminiscent of typical experiments. Different overlaps between the electronic states of the leads and the systems' orbitals generally lead to a situation where $\Gamma_1 \neq \Gamma_2$. Now, we further assume a small finite hopping between the lower orbitals leading to an additional Hamiltonian term $- t_1 \sum_{j=1}^{N-1} \left(c^{\dagger}_{1,j+1} c_{1,j} + \rm{h.c.} \right)$, with $t_1 \ll \Gamma_1$. We choose $t_{2} \gg t_{1}$ as spatially extended upper orbitals generally exhibit larger overlaps than valence orbitals.  In an asymmetric situation, where one has a poor injection/extraction rate in the upper band with large hopping, and vice versa for the lower band, photon dressing of the two bands then allows for dramatic current enhancement. Now, the total current has two contributions even for $g=0$, and still considering $t_2 \gg \Gamma_2$, Eq.~\hyperref[eq:stb]{(\ref*{eq:stb})} is extended to $J = ({e \Gamma_1}/{2}) [ ({2 t_1}/{\Gamma_1})^2  + {\Gamma_2}/{\Gamma_1}]$. While previously the second term was $\Gamma_2/\Gamma_1 = 1$, now $\Gamma_2 / \Gamma_1 \ll 1$, and for $g\neq 0$, the current restored in the lower band becomes the dominant contribution. Ultimately, when $t_1 \ll \Gamma_1$, the relative current enhancement is only limited by the ratio $\Gamma_1/\Gamma_2$ \cite{SM}. We note that the limit $\Gamma_2 \to 0$ is equivalent to the low-bias regime (when the Fermi level at zero-voltage lies in the lower band), where electrons can only be injected and extracted in the lower band. 

In a plausible scenario, electrons will also be injected into defect states which in turn can spontaneously decay and provide residual population in the photon bath. This residual population has a considerable effect, as e.g.~for $\Gamma_2 \to 0$, it is needed to initiate the effective hopping process through the dressed bands. In Fig.~\ref{fig:phot}, we show the current enhancement for mean photon population in the bath $N_{\rm ph}=0,0.5$ and $\Gamma_2 \ll \Gamma_1$. Panels {\bf (a)} and {\bf (b)} display the individual and collective dressing regimes, respectively. Both panels demonstrate order-of-magnitude enhancements even for $N_{\rm ph} \leq 1$, i.e.~a cavity mode close to the vacuum. In both cases, increasing the photon population boosts the current significantly further. In \cite{SM}, we show that the latter originates from a bosonic enhancement scaling as $g \sqrt{N_{\rm ph}}$.

% CONCLUSION & OUTLOOK
In this work, we introduced a proof-of-principle mechanism to enhance charge conductivity in a mesoscopic chain by coupling it to the vacuum field of a cavity. We showed that this enhancement can reach orders of magnitude. It is an exciting prospect to explore how this model can be extended to explain recent experiments with two-dimensional organic semiconductors, where large enhancements of charge conductivity have been reported~\cite{orgiu2015conductivity}. Note that we expect our findings to qualitatively hold in the case of a direct generalization of our model to higher dimensions. Our model might also find applications in several other fields, such as nanowires \cite{rurali_colloquium_2010}, carbon nanotubes \cite{laird_quantum_2015} or quantum dot arrays \cite{wang_fabrication_2004,kagan_charge_2015}. In particular, pairs of quantum dots have recently been coupled to microwave cavities \cite{liu_semiconductor_2015, frey_dipole_2012,viennot_out--equilibrium_2014}. Possible extensions of our model include coupling to multiple transmission channels in different geometries, as well as the competition between light-matter coupling and Anderson localization in random lattices. In general, the method used in this article provides new perspectives for the investigation of many-body systems strongly coupled to cavity resonances or other bosonic (e.g.~phononic) degrees of freedom.

\paragraph{Acknowledgements} We are grateful to Thibault Chervy, Roberta Citro, Thomas Ebbesen, Cyriaque Genet, Emanuele Orgiu, and Paolo Samor\`{i} for inspiring discussions. Work in Strasbourg was supported by the ERC St-Grant ColDSIM (No. 307688), with additional funding from Rysq and ANR-FWF grant BLUESHIELD. C.G. acknowledges support from the Max Planck Society and from the COST action NQO 1403 (Nano-scale Quantum Optics).
					
\bibliography{charge}

%\pagebreak
\clearpage

\widetext

\section{Supplemental Material}

\subsection{Steady-state current and Green's functions} 
\label{ss_cucu_sec}

In this section, we present the non-equilibrium Green's function (NEGF) method, based on a generalization of the model presented in the Chapter 12 of Ref.~[\onlinecite{haug}]. In the following, we will consider $\hbar=1$, and use the short-hand notations $\partial_{\tau}\equiv \frac{\partial}{\partial \tau}$ and $\delta_{f (\tau)}\equiv \frac{\delta}{\delta f(\tau)}$, for function and functional derivatives respectively. For sake of generality, we will consider two different nearest-neighbors hopping rates $t_{1}$ and $t_{2}$ between the lower and upper orbitals respectively, as well as different injection/extraction rates $\Gamma_{1}$ and $\Gamma_{2}$. The results discussed in the main text for $\Gamma_{1}=\Gamma_{2}\equiv \Gamma$, $t_{1}=0$ and $t_{2}\equiv t$ can be recovered as a particular case of the general method presented here. Formally including the counter-rotating terms in the coupling Hamiltonian, the Hamiltonian which describes the mesoscopic chain is $H_{S}=H_{\rm el} + H_{\rm int}+H_{\rm cav}$, where

\begin{align}
&H_{\rm el} = \sum_{j=1}^{N} \sum_{\alpha=1}^{2} \omega_{\alpha} c^{\dagger}_{\alpha,j} c_{\alpha,j} - \sum_{\alpha=1}^{2} t_{\alpha} \sum_{j=1}^{N-1} \left(c^{\dagger}_{\alpha,j+1} c_{\alpha,j} + \rm{h.c.} \right) \label{H_kinetic} \\
&H_{\rm int} = g \sum_{j=1}^{N} \left(c^{\dagger}_{2,j} c_{1,j} + c^{\dagger}_{1,j} c_{2,j} \right) A
\label{H_intera} \\
&H_{\rm cav}=\omega_{0} a^\dagger a. 
\label{H_bos0}
\end{align}

Here, $A=a+a^{\dagger}$, with $a$ and $a^{\dagger}$ the bosonic annihilation and creation operators for the cavity mode with energy $\omega_{0}$. The fermionic operator $c_{\alpha,j}$ ($c^{\dagger}_{\alpha,j}$) annihilates (creates) an electron in the orbital $\alpha=1,2$ on site $j$. Denoting by $\eta=s$ and $\eta=d$, the source and the drain leads, the coupling of the chain to the leads can be described by the tunnelling Hamiltonian:

\begin{align}
H_{T} &=\sum_{\alpha}\sum_{\eta=s,d} \sum_{\bf q} \omega_{\bf q} b^{\dagger}_{\alpha,{\bf q},\eta} b_{\alpha,{\bf q},\eta} + \sum_{\alpha} \sum_{\eta=s,d} \sum_{j,{\bf q}}  \lambda_{\alpha,j,{\bf q},\eta} \left( c_{\alpha,j} b^{\dagger}_{\alpha,{\bf q},\eta} + b_{\alpha,{\bf q},\eta} c^{\dagger}_{\alpha,j}\right), 
\end{align}

with coupling constants $\lambda_{\alpha,j,{\bf q},s} = \lambda_{\alpha,{\bf q}}$ for $j=1$, $\lambda_{\alpha,j,{\bf q},s} = 0$ for $j\neq 1$, and $\lambda_{\alpha,j,{\bf q},d} = \lambda_{\alpha,{\bf q}}$ for $j=N$, $\lambda_{\alpha,j,{\bf q},d} = 0$ for $j\neq N$. The operators $b^{\dagger}_{\alpha,{\bf q},\eta}$ ($b_{\alpha,{\bf q},\eta}$) create (annihilate) a fermion in the state $(\alpha,{\bf q})$ of the lead $\eta$, and obey fermionic commutation relations. ${\bf q}$ is a continuous index interpreted as the 2d electron wavevector in the leads. The cavity mode is coupled to a photonic bath described by a 3d mode continuum with wavevectors ${\bf p}$. The corresponding Hamiltonian is written in terms of the extra-cavity photon operators $a^{\dagger}_{\bf p}$ and $a_{\bf p}$ (obeying bosonic commutation rules) as:

\begin{equation}
H_{\rm ph} = \sum_{\bf p} \omega_{\bf p} a^{\dagger}_{\bf p} a_{\bf p} + \sum_{\bf p} \mu_{\bf p} A_{\bf p} A,
\end{equation}

where $A_{\bf p}=a_{\bf p}+a^{\dagger}_{\bf p}$. Using Fourier series of the chain operators $c_{\alpha,j}=\sum_{k=1}^{N} \varphi^{j}_{k} \tilde{c}_{\alpha,k}$, with $\varphi^{j}_{k}=\sqrt{2/(N+1)} \sin \left(\pi j k/(N+1) \right)$, the total (system+reservoirs) Hamiltonian $H=H_{S}+H_{\rm ph}+H_{T}$ can be written in $k$-space as: 

\begin{align}
H_{S} &= \sum_{\alpha,k} \omega_{\alpha,k} \tilde{c}^{\dagger}_{\alpha,k} \tilde{c}_{\alpha,k} + \omega_{0} a^{\dagger} a + g \sum_{k} \left(\tilde{c}^{\dagger}_{2,k} \tilde{c}_{1,k} + \tilde{c}^{\dagger}_{1,k} \tilde{c}_{2,k} \right) A  \nonumber \\
H_{T} &=  \sum_{\alpha}\sum_{\eta,{\bf q}} \omega_{\bf q} b^{\dagger}_{\alpha,{\bf q},\eta} b_{\alpha,{\bf q},\eta} + \sum_{\eta,{\bf q}} \sum_{\alpha,k} \lambda_{\alpha,{\bf q}} \varphi^{j_{\eta}}_{k} \left( \tilde{c}_{\alpha,k} b^{\dagger}_{\alpha,{\bf q},\eta} + b_{\alpha,{\bf q},\eta} \tilde{c}^{\dagger}_{\alpha,k} \right),
\label{hh_toto}
\end{align}

where $\omega_{\alpha,k}=\omega_{\alpha}-2t_{\alpha}\cos(\pi k/(N+1))$, $j_{s}=1$ and $j_{d}=N$. In the steady-state, the charge current $J_{\eta}$ flowing through the lead $\eta$ is such that $J_{s}=-J_{d}$, and is given by $J_{\eta} = -e \partial_t \langle N_{\eta} \rangle = - i e \langle [H,N_{\eta}] \rangle$, where $\langle \cdots \rangle$ stands for quantum average in the ground state of the total Hamiltonian $H$, and $N_{\eta}=\sum_{\alpha,{\bf q}} b^{\dagger}_{\alpha,{\bf q},\eta} b_{\alpha,{\bf q},\eta}$ is the number of electrons in the lead $\eta$. A direct calculation of the commutator allows to first express the current in terms of a Green's function (GF) which describes the correlations between the leads and the atom chain:

\begin{equation}
J_{\eta} = -2e\sum_{\alpha,k} \sum_{\bf q} \varphi^{j_{\eta}}_{k} \lambda_{\alpha,{\bf q}} \int \!\! \frac{d\omega}{2\pi} \Re \left[ G^{<}_{\alpha,k,{\bf q},\eta} (\omega) \right], 
\label{current_2_re}
\end{equation} 

where $\Re$ stands for real part, and $G^{<}_{\alpha,k,{\bf q},\eta} (\omega)=i \int \! d(\tau-\tau') e^{i \omega (\tau-\tau')} \langle b^{\dagger}_{\alpha,{\bf q},\eta} (\tau') \tilde{c}_{\alpha,k} (\tau) \rangle$. Introducing the time-ordered mixed system-leads GF $G_{\alpha,k,{\bf q},\eta} (\tau-\tau')= -i \langle \mathcal{T} \tilde{c}_{\alpha,k} (\tau) b^{\dagger}_{\alpha,{\bf q},\eta} (\tau') \rangle$, where the time-ordered product is defined as $\mathcal{T} A(\tau) B(\tau')=\Theta (\tau-\tau')A(\tau) B(\tau') - \Theta (\tau'-\tau) B(\tau') A(\tau)$, the equation of motion satisfied by $G_{\alpha,k,{\bf q},\eta}$ can be found by taking the time derivative $\partial_{\tau'} G_{\alpha,k,{\bf q},\eta} (\tau-\tau')$, and then computing the different commutators that enter the Heisenberg equation $\partial_{\tau'} b^{\dagger}_{\alpha,{\bf q},\eta} (\tau')=i[H,b^{\dagger}_{\alpha,{\bf q},\eta}] (\tau')$:

\begin{align}
\left(- i \frac{\partial}{\partial \tau'}  - \omega_{\bf q} \right) G_{\alpha,k,{\bf q},\eta}(\tau-\tau') = - \lambda_{\alpha,{\bf q}} \sum_{k'} \varphi^{j_{\eta}}_{k'}  G_{\alpha,k,k'}(\tau-\tau').
\end{align}

This equation of motion is then solved as:

\begin{equation}
G_{\alpha,k,{\bf q},\eta} (\omega)= - \lambda_{\alpha,{\bf q}} \sum_{k'} \varphi^{j_{\eta}}_{k'} G_{\alpha,k,k'} (\omega) \mathcal{G}_{{\bf q},\eta} (\omega),
\label{final_GF_mixed00}
\end{equation}

where $G_{\alpha,k,k'} (\omega) = -i \int \! d(\tau-\tau') e^{i \omega (\tau-\tau')} \langle \mathcal{T} \tilde{c}_{\alpha,k} (\tau) \tilde{c}^{\dagger}_{\alpha,k'} (\tau') \rangle$ denotes the time-ordered GF of the chain, and the leads time-ordered GFs are defined as $\mathcal{G}_{{\bf q},\eta} (\omega)=-i \int \! d(\tau-\tau') e^{i \omega (\tau-\tau')} \langle \mathcal{T} b_{\alpha,{\bf q},\eta} (\tau) b^{\dagger}_{\alpha,{\bf q},\eta} (\tau') \rangle_{0}=1/(\omega-\omega_{\bf q})$. Here, $\langle \cdots \rangle_{0}$ refers to the expectation value in the ground state of the lead Hamiltonian $\sum_{\alpha,\eta,{\bf q}} \omega_{\bf q} b^{\dagger}_{\alpha,{\bf q},\eta} b_{\alpha,{\bf q},\eta}$. We now use the Langreth rules~\cite{haug} on Eq.~(\ref{final_GF_mixed00}), and get:

\begin{align}
G^{<}_{\alpha,k,{\bf q},\eta} (\omega) =- \lambda_{\alpha,{\bf q}} \sum_{k'} \varphi^{j_{\eta}}_{k'} \Big(G^{r}_{\alpha,k,k'} (\omega) \mathcal{G}^{<}_{{\bf q},\eta} (\omega) + G^{<}_{\alpha,k,k'} (\omega) \mathcal{G}^{a}_{{\bf q},\eta} (\omega)\Big),
\label{gren_lead_el}
\end{align}

where $r$ and $a$ stand for retarded and advanced GFs respectively. Converting the summation over ${\bf q}$ in Eq.~(\ref{current_2_re}) into a frequency integral $\sum_{\bf q} \to \int_{0}^{\infty} d\omega \rho (\omega)$, where $\rho (\omega)$ represents the electron density of states in the leads, we introduce the tunnelling rate between the chain and the leads as $\Gamma_{\alpha} (\omega)=2\pi \rho (\omega) \lambda^{2}_{\alpha} (\omega)$. For sake of simplicity, we ignore the energy dependence of the tunnelling rate ($\Gamma_{\alpha} (\omega) \equiv \Gamma_{\alpha}$), which amounts to neglect the non-local temporal response (Markovian leads). Using the results: 

\begin{align}
&\mathcal{G}^{<}_{{\bf q},\eta} (\omega) = 2i\pi \delta (\omega - \omega_{\bf q} ) n_{\eta} (\omega) \quad \textrm{and} \quad \mathcal{G}^{a}_{{\bf q},\eta} (\omega) = \frac{1}{\omega - \omega_{\bf q} - i 0^{+}},&
\label{gg_petites}
\end{align}

where $0^{+}$ an infinitesimal positive quantity, and $n_{\eta} (\omega)$ is the Fermi occupation number of the lead $\eta$, and considering the high-bias regime, i.e. $n_{s} (\omega)=1$ and $n_{d} (\omega)=0$, the steady-state current can be put on the form:

\begin{equation}
J = \frac{J_{s}-J_{d}}{2} =\sum_{\alpha} \frac{e \Gamma_{\alpha}}{2} \int \!\! \frac{d\omega}{2\pi} T_{\alpha}(\omega),
\label{transsm}
\end{equation}

where the transmission coefficient is derived as:

\begin{align}
T_{\alpha}(\omega) &= \textbf{Tr} \left[-2\underline{\sigma}^{1} \circ \Im \underline{G}^{r}_{\alpha} (\omega) + \left( \underline{\sigma}^{N} -\underline{\sigma}^{1}\right) \circ \Im  \underline{G}^{<}_{\alpha} (\omega) \right].
\label{trans_long_coco}
\end{align}

Here $\textbf{Tr} \equiv \sum_{k,k'}$, underlined quantities denote $N\times N$ matrices, $\circ$ is the element-wise Hadamard product, and $\Im$ stands for imaginary part. The matrix elements of $\underline{\sigma}^{j}$ are given by $\sigma^{j}_{k,k'}=\varphi^{j}_{k}\varphi^{j}_{k'}$. 

\subsection{Electron Green's functions} 
\label{ee_GF_funsec}

Let us show that the electron GFs in the chain satisfy a Dyson equation of motion. As before, we first compute the time derivative $\partial_{\tau} G_{\alpha,k,k'} (\tau-\tau')$ combined with the Heisenberg equation $\partial_{\tau} \tilde{c}_{\alpha,k} (\tau)=i[H,\tilde{c}_{\alpha,k}] (\tau)$, and obtain: 

\begin{align}
\left( i \partial_\tau - \omega_{\alpha,k} \right) G_{\alpha,k,k'} (\tau-\tau')= \delta_{k,k'} \delta (\tau-\tau') + g (1-\delta_{\alpha,\alpha'}) F_{\alpha',k,\alpha,k'} (\tau-\tau') - \sum_{{\bf q},\eta} \lambda_{\alpha,{\bf q}} \varphi^{j_{\eta}}_{k} G_{{\bf q},\eta,\alpha,k'}(\tau-\tau'),
\label{eq_momo_long_nc}
\end{align} 

where $G_{{\bf q},\eta,\alpha,k} (\tau-\tau')= -i \langle \mathcal{T} b_{\alpha,{\bf q},\eta} (\tau) \tilde{c}^{\dagger}_{\alpha,k} (\tau') \rangle$ is a mixed system-leads GF, and $F_{\alpha',k,\alpha,k'} (\tau-\tau') = -i \langle \mathcal{T} \tilde{c}_{\alpha',k} (\tau) \tilde{c}^{\dagger}_{\alpha,k'} (\tau') A (\tau) \rangle$ corresponds to a higher-order correlation function mixing the electronic and photonic degrees of freedom. The equation of motion for $G_{{\bf q},\eta,\alpha,k}$ can be derived similarly as in the previous section, providing:

\begin{align}
G_{{\bf q},\eta,\alpha,k'} (\omega) &= - \lambda_{\alpha,{\bf q}} \sum_{k_{1}} \varphi^{j_{\eta}}_{k_{1}} \mathcal{G}_{{\bf q},\eta} (\omega) G_{\alpha,k_{1},k'}(\omega).
\label{eq_of_motion2121} 
\end{align}

Following the procedure of Ref.~[\onlinecite{engelsberg1963coupled}], the correlation function $F_{\alpha',k,\alpha,k'} (\tau-\tau')$ can now be written in terms of single-particle GFs, by adding a vanishing source term $H'= \mathcal{J} A$ to the Hamiltonian $H$, and then taking the functional derivative $\delta_{\mathcal{J} (\tau)} G_{\alpha',k,\alpha,k'} (\tau-\tau')$, with:

\begin{equation}
G_{\alpha',k,\alpha,k'} (\tau-\tau') = -i \frac{\langle \mathcal{T} \tilde{c}_{\alpha',k} (\tau) \tilde{c}^{\dagger}_{\alpha,k'} (\tau') e^{-i \int \! d \tau_{1} H (\tau_{1})}\rangle_{0}}{\langle e^{-i \int \! d \tau_{1} H (\tau_{1})} \rangle_{0}}. 
\end{equation} 
 
The operation $\langle \cdots \rangle_{0}$ stands for quantum average in the ground state of the non-interacting Hamiltonian $H_{0}=H_{\rm el}+H_{\rm cav}$ defined in the main text. After some manipulations using the functional derivative properties, we obtain:

\begin{align}
F_{\alpha',k,\alpha,k'} (\tau-\tau') = i g \sum_{k_{1},k_{2}} \int \!\! d\tau_{1} \!\! \int \!\! d\tau_{2} \!\! \int \!\! d\tau_{3} G_{\alpha',k,k_{1}} (\tau-\tau_{1}) \Lambda_{\alpha',k_{1},\alpha,k_{2}} (\tau_{1},\tau_{2},\tau_{3}) D (\tau_{3} - \tau) G_{\alpha,k_{2},k'} (\tau_{2}-\tau'),
\label{ff_fun_app}
\end{align} 

where the time-ordered photon Green function is defined as $D (\tau_{3} - \tau)=\delta_{\mathcal{J} (\tau)} \langle A (\tau_{3})\rangle = -i \langle \mathcal{T} A ( \tau_{3}) A (\tau) \rangle$, and the so-called vertex function $\Lambda_{\alpha',k_{1},\alpha,k_{2}} (\tau_{1},\tau_{2},\tau_{3}) = -\frac{1}{g} \delta_{\langle A (\tau_{3})\rangle} G^{-1}_{\alpha',k_{1},\alpha,k_{2}} (\tau_{1}-\tau_{2})$. It can be shown that this vertex function satisfies a self-consistent equation\cite{engelsberg1963coupled} of the kind: 

\begin{align}
\Lambda_{\alpha',k_{1},\alpha,k_{2}} (\tau_{1},\tau_{2},\tau_{3}) = \left(1-\delta_{\alpha',\alpha} \right) \delta_{k_{1},k_{2}} \delta( \tau_{1}-\tau_{2}) \delta (\tau_{1} - \tau_{3}) + \mathcal{O} (\Lambda^{2}),
\label{vert_corr1}
\end{align}

where $\mathcal{O} (\Lambda^{2})$ denotes a second order functional in $\Lambda$, providing vertex corrections\cite{engelsberg1963coupled} (crossed diagrams). In the self-consistent Born approximation (SCBA), vertex corrections are neglected and we keep only the first term in the right-hand side of Eq.~(\ref{vert_corr1}). Back to Eq.~(\ref{eq_momo_long_nc}), we convert the summation over ${\bf q}$ into a frequency integral (see Sec.~\ref{ss_cucu_sec}), and then use Eqs.~(\ref{eq_of_motion2121}),~(\ref{ff_fun_app}), and~(\ref{vert_corr1}). Equation~(\ref{eq_momo_long_nc}) finally takes a Dyson form in the frequency domain:

\begin{align}
\sum_{k_{1}} \left((G^{0}_{\alpha,k,k_{1}} (\omega))^{-1} - \Sigma_{\alpha,k,k_{1}} (\omega) \right) G_{\alpha,k_{1},k'} (\omega) =\delta_{k,k'},
\label{dyson_aspi}
\end{align}

with the SCBA self-energy (SE): 

\begin{align}
\Sigma_{\alpha,k,k'} (\omega) &= i g^{2} \left(1-\delta_{\alpha_{1},\alpha}\right) \int \!\! \frac{d\omega'}{2\pi} G_{\alpha_{1},k,k'} (\omega+\omega') D (\omega') + \sum_{{\bf q},\eta} \lambda^{2}_{\alpha,{\bf q}} \varphi^{j_{\eta}}_{k} \varphi^{j_{\eta}}_{k'} \mathcal{G}_{{\bf q},\eta} (\omega),
\label{lm_se_lala36}
\end{align}  

and the non-interacting GF $G^{0}_{\alpha,k,k_{1}} (\omega)=\delta_{k,k_{1}}/(\omega - \omega_{\alpha,k})$. Still in the high-bias regime, one can then use the Langreth rules together with Eq.~(\ref{gg_petites}), to find the ``lesser'' SE:

\begin{align}
\underline{\Sigma}^{<}_{\alpha} (\omega)= i g^{2} \left(1-\delta_{\alpha,\alpha'} \right)\int \!\! \frac{d\omega'}{2\pi} \underline{G}^{<}_{\alpha'} (\omega+\omega') D^{>} (\omega') + i\Gamma_{\alpha} \underline{\sigma}^{1},
\label{lm_se_lala}
\end{align}

where $D^{>}(\omega)$ denotes the ``greater'' photon GF. The first contribution of Eq.~(\ref{lm_se_lala}) represents the electron SE correction due to the light-matter coupling, stemming from the emission/absorption of cavity excitations (i.e.~the poles of $D(\omega)$) when electrons undergo optical transitions between the two bands. The second term $\propto \Gamma_{\alpha}$ is exact, and represents the broadening of electron states due to the coupling between the  chain and the leads. It is worth mentioning that the validity domains of the SCBA and the rotating wave-approximation~\cite{scully} [used in Eq.~(3) of the main text] overlap in such a way that the SCBA would break down if the contribution due to the counter rotating terms $\propto (c^{\dagger}_{2,j} c_{1,j} a^{\dagger}+\textrm{h.c.})$ in Eq.~(\ref{H_intera}) becomes important (when $g/\omega_{21} \lesssim 1$). 

The ``greater'' SE $\underline{\Sigma}^{>}_{\alpha} (\omega)$ exhibits a similar expression as Eq.~(\ref{lm_se_lala}). From there, one can either use the Langreth rules on Eq.~(\ref{lm_se_lala36}) to calculate the retarded and advanced SEs, or proceed as explained in Sec.~\ref{self_cons_proc}. The retarded and advanced GFs can then be computed from the Dyson equation (\ref{dyson_aspi}): $\underline{G}^{\beta}_{\alpha} (\omega) = [(\underline{G}^{0\beta}_{\alpha}(\omega))^{-1} - \underline{\Sigma}^{\beta}_{\alpha} (\omega)]^{-1}$ ($\beta=r,a$). On the other hand, $\underline{G}^{<}_{\alpha} (\omega)$ can be calculated through the Keldysh equation $\underline{G}^{<}_{\alpha} (\omega)=\underline{G}^{r}_{\alpha} (\omega) \underline{\Sigma}^{<}_{\alpha} (\omega) \underline{G}^{a}_{\alpha} (\omega)$.

\subsection{Cavity photons Green's functions}
\label{ph_sec_app}

Let us now show that the time-ordered photon GF $D(\omega)$ satisfies a Dyson equation. We start by taking the second time derivative of the cavity vector potential $A(t)$, and then use the Heisenberg equation $\partial_{\tau} A (\tau)=i [H,A ] (\tau)$ two times. As in the previous section, we consider a vanishing source term $H'= \mathcal{J} A$ in the Hamiltonian $H$. Then, we take the functional derivative of the ground-state expectation of the obtained equation with respect to $\mathcal{J} (\tau')$. This calculation yields the following equation of motion for $D(\tau-\tau')$:

\begin{align}
\left(- \frac{\partial^{2}_{\tau}}{2\omega_{0}} - \frac{\omega_{0}}{2} \right) D (\tau-\tau') &=  \delta (\tau - \tau') - i g \sum_{\alpha,\alpha'} \sum_{k} \left(1-\delta_{\alpha,\alpha'} \right) \delta_{\mathcal{J} (\tau')} G_{\alpha,k,\alpha',k} (\tau,\tau^{+}) + \sum_{\bf p} \mu_{\bf p} D_{\bf p} (\tau-\tau'),
\label{eq_momo_ph}
\end{align} 

where the time $\tau^{+}$ is $\tau$ plus a positive vanishing quantity, and $D_{\bf p} (\tau-\tau')= -i \langle \mathcal{T} A_{\bf p} (\tau) A (\tau') \rangle$ describes the correlations between the cavity mode and the electromagnetic environment. The equation of motion satisfied by $D_{\bf p}$ can be derived similarly as before (by taking its second time derivative), namely: 

\begin{align}
\left(-\partial^{2}_{\tau} - \omega^{2}_{\bf p}\right) D_{\bf p} (\tau-\tau') = 2 \omega_{\bf p} \mu_{\bf p}  D (\tau-\tau'),
\label{equ_mot_ph_app}
\end{align} 

which is formally solved in the frequency domain as $D_{\bf p} (\omega) = \mu_{\bf p}\mathcal{D}_{\bf p} (\omega) D (\omega)$. Here, ${\mathcal{D}_{\bf p} (\omega)= -i \int \! d(\tau-\tau') e^{i \omega (\tau-\tau')} \langle \mathcal{T} A_{\bf p} (\tau) A_{\bf -p} (\tau') \rangle_{0}=2\omega_{\bf p}/(\omega^{2}-\omega^{2}_{\bf p})}$ denotes the extra-cavity photon GF. On the other hand, the second term in the right-hand side of Eq.~(\ref{eq_momo_ph}) can be calculated in a similar fashion as in Sec.~\ref{ee_GF_funsec}, namely:

\begin{align}
\delta_{\mathcal{J} (\tau')} G_{\alpha,k,\alpha',k} (\tau,\tau^{+}) =  g \sum_{k_{1},k_{2}} \int \!\! d\tau_{1} \int \!\! d\tau_{2} \int \!\! d\tau_{3} G_{\alpha,k,k_{1}} (\tau-\tau_{1}) \Lambda_{\alpha,k_{1},\alpha',k_{2}} (\tau_{1},\tau_{2},\tau_{3})  D (\tau_{3} - \tau') G_{\alpha',k_{2},k} (\tau_{2}-\tau^{+}),
\label{green_12}
\end{align} 

where the vertex function is defined in Eq.~(\ref{vert_corr1}). In the SCBA, we assume $\Lambda_{\alpha,k_{1},\alpha',k_{2}} (\tau_{1},\tau_{2},\tau_{3})=\left(1-\delta_{\alpha,\alpha'} \right) \delta_{k_{1},k_{2}} \delta(\tau_{1}-\tau_{2}) \delta (\tau_{1} - \tau_{3})$, which inserted in Eq.~(\ref{green_12}), allows to write the equation of motion (\ref{eq_momo_ph}) in the Dyson form $\left(D^{-1}_{0} (\omega) - \Pi (\omega) \right) D (\omega) = 1$, with the bare photon GF $D_{0}(\omega)=2\omega_{0}/(\omega^{2} - \omega^{2}_{0})$, and the photon SE:

\begin{align}
\Pi (\omega) &= -i g^{2} \sum_{\alpha,\alpha'} \sum_{k,k_{1}} \left(1-\delta_{\alpha,\alpha'} \right) \int \!\! \frac{d\omega'}{2\pi} G_{\alpha,k,k_{1}} (\omega+\omega') G_{\alpha',k_{1},k} (\omega') + \sum_{\bf p} \mu^{2}_{\bf p} \mathcal{D}_{\bf p} (\omega).
\label{ph_se_2}
\end{align}

The summation over ${\bf p}$ is then converted into a frequency integral, $\sum_{\bf p} \to \int_{0}^{\infty} d\omega \rho_{0} (\omega)$, where $\rho_{0} (\omega)$ denotes the extra-cavity photon density of states. We introduce the cavity photon decay rate as $\kappa (\omega) = 2\pi \rho_{0} (\omega) \mu^{2} (\omega)$ (assumed to be frequency-independent, i.e. $\kappa (\omega) \equiv \kappa$), and use the Langreth rules in Eq.~(\ref{ph_se_2}). Introducing the mean population in the photon bath as $N_{\rm ph}=\langle a^{\dagger}_{\bf p} a_{\bf p}\rangle$ (also assumed to be frequency-independent), we get $\mathcal{D}^{>}_{\bf p} (\omega)=-2i\pi (N_{\rm ph}+1)\delta(\omega-\omega_{\bf p})-2i\pi N_{\rm ph} \delta(\omega+\omega_{\bf p})$, and the ``greater'' photon SE can thus be written as:

\begin{align}
\Pi^{>} (\omega)=-i g^{2}\left(1-\delta_{\alpha,\alpha'} \right) \textbf{Tr} \int \!\! \frac{d\omega'}{2\pi} \underline{G}^{>}_{\alpha} (\omega+\omega')\underline{G}^{<}_{\alpha'} (\omega') -i\kappa \left((N_{\rm ph}+1) \theta (\omega) + N_{\rm ph} \theta (-\omega)\right),
\label{popo_ph_less}
\end{align}	

where the product of the two electron GFs is a matrix multiplication, $\theta$ is the Heaviside function, and $\textbf{Tr} \equiv \sum_{\alpha,k,k'}$. The first contribution can be identified with the polarization function associated with the transition dipole moments, and provides a dressing of the bare cavity photon GF $D_{0}$. The second contribution $\propto \kappa$ describes the coupling between the cavity mode and the photon bath, and is calculated exactly.

Similarly to electrons, the ``greater'', retarded, and advanced photon SEs can be computed by using the Langreth rules on Eq.~(\ref{ph_se_2}), or alternatively by following the procedure given in Sec.~\ref{self_cons_proc}. The ``greater'' photon GF can then be calculated using the Keldysh equation $D^{>} (\omega)=D^{r}(\omega) \Pi^{>}(\omega) D^{a}(\omega)$, where retarded and advanced photon GFs are given by the Dyson equation $D^{\beta} (\omega) = [(D^{\beta}_{0} (\omega))^{-1} - \Pi^{\beta} (\omega)]^{-1}$, with $\beta=r,a$. Furthermore, the retarded photon GF can be used to define the normalized cavity photon DOS as $A_{c}(\omega)=-2\Im D^{r}(\omega)$, which can be directly accessed experimentally by measuring the cavity absorption spectrum. 
 
Note that the photon GF $D^{<} (\omega)$ is directly related to the mean cavity photon number in the steady state (up to small squeezing terms) as $N_{\rm cav} = -\frac{1}{2} \left(\int \frac{d\omega}{2\pi} \Im D^{<}(\omega) +1 \right)$. As expected, it can be checked numerically that the mean population in the cavity and in the photon bath coincide in the steady-state, namely $N_{\rm cav} \approx N_{\rm ph}$.  

\subsection{Self-consistent procedure}
\label{self_cons_proc}

In the previous sections, we have shown that electron/photon SEs and GFs are related to each other by a self-consistent, closed set of integro-differential equations. The numerical procedure to solve these equations self-consistently is as follow: we start with the electron SEs $\underline{\Sigma}^{\lessgtr}_{\alpha}$ defined in Eq.~(\ref{lm_se_lala}). The first order is obtained by replacing the fully interacting electron GF $\underline{G}^{\lessgtr}_{\alpha}$ by the non-interacting one $\underline{G}^{0\lessgtr}_{\alpha}$, evaluated in the ground state of $H_{0}$: 

\begin{align}
&G^{0<}_{\alpha,k,k'} (\omega)= 2i\pi n^{0}_{\alpha,k} \delta_{k,k'} \delta(\omega-\omega_{\alpha,k}), \quad G^{0>}_{\alpha,k,k'} (\omega)= -2i\pi (1-n^{0}_{\alpha,k}) \delta_{k,k'} \delta(\omega-\omega_{\alpha,k}),&
\end{align}

with $n^{0}_{\alpha,k}$ the Fermi occupation number of the Bloch state $k$ in the band $\alpha'$, for $g=0$. Similarly, $D^{\lessgtr}$ are replaced by $D^{\lessgtr}_{1}$, calculated without light-matter coupling ($g=0$) but in the presence of the photon bath [second term in Eq.~(\ref{popo_ph_less})]. Considering the resonant case $\omega_{0}=\omega_{21} \equiv 1$, one finds:

\begin{align}
D^{>}_{1}(\omega)= \frac{-4i \kappa}{\left(\omega^{2}-1\right)^{2}+\kappa^{2}} \Big((N_{\rm ph}+1) \theta (\omega) + N_{\rm ph} \theta (-\omega)\Big).
\label{D1_less}
\end{align}

The retarded and advanced electron SEs can then be efficiently calculated using the real-time equality\cite{pourfath} $\underline{\Sigma}^{r}_{\alpha} (t)=\theta(t) \left(\underline{\Sigma}^{>}_{\alpha} (t) - \underline{\Sigma}^{<}_{\alpha} (t) \right)$. Introducing the real function $\underline{\chi}_{\alpha} (\omega)=i\left(\underline{\Sigma}^{>}_{\alpha} (\omega) - \underline{\Sigma}^{<}_{\alpha} (\omega) \right)$, the previous equality can be written in the frequency domain as:

\begin{align} 
\underline{\Sigma}^{r}_{\alpha} (\omega)= -i\frac{\underline{\chi}_{\alpha} (\omega)}{2} + \frac{1}{2\pi} \textrm{p.v} \int d\omega' \! \frac{\underline{\chi}_{\alpha} (\omega')}{\omega-\omega'},
\label{rere_self_causal} 
\end{align} 

where p.v denotes the Cauchy principal value. As a causal function, the real and imaginary parts of $\underline{\Sigma}^{r}_{\alpha} (\omega)$ are related to each other by Kramers-Kronig relations, as it can be checked directly from Eq.~(\ref{rere_self_causal}). The advanced SE is simply given by $\underline{\Sigma}^{a}_{\alpha} (\omega)=\left(\underline{\Sigma}^{r}_{\alpha} (\omega)\right)^{\dagger}$. The function $\underline{\chi}_{\alpha} (\omega)$ is related to the spectral broadening of Bloch states induced by the coupling to the leads and the cavity mode, while the real part of $\underline{\Sigma}^{r}_{\alpha} (\omega)$ provides a shift of the Bloch state energies $\omega_{\alpha,k}$. Plugging these results in the Dyson and Keldysh equations, we now calculate the first-order electron GFs $\underline{G}^{\beta}_{\alpha}$ ($\beta=r,a$) and $\underline{G}^{\lessgtr}_{\alpha}$, and use the latter to compute the first-order photon SEs $\Pi^{\lessgtr}$ defined in Eq.~(\ref{popo_ph_less}). The retarded and advanced parts $\Pi^{\beta}$ are obtained using the relation $\Pi^{r} (t)=\theta(t) \left(\Pi^{>} (t) - \Pi^{<} (t) \right)$, and can be inserted into the Dyson and Keldysh equations to calculate the first-order photon GFs $D^{\beta}$ and $D^{\lessgtr}$. Finally, the second-order electron SE $\underline{\Sigma}^{\lessgtr}_{\alpha}$ is obtained from Eq.~(\ref{lm_se_lala}), with $D^{\lessgtr}$ and $\underline{G}^{\lessgtr}_{\alpha}$, and the whole cycle is then repeated until convergence. Namely, this procedure is repeated a sufficient number of times, such that the total current flowing through the bands has converged. 

\subsection{Current in the absence of light-matter coupling}
\label{cucu_g_spec}

The total current flowing through the chain in the absence of light-matter coupling ($g=0$) can be calculated using the results of the previous sections. The current $J_{\alpha}$ flowing through the band $\alpha$ is given by Eqs.~(\ref{transsm}) and (\ref{trans_long_coco}). Using the spectral function sum rule $\int \! d\omega A_{\alpha,k,k'}(\omega) =2\pi \delta_{k,k'}$, with $A_{\alpha,k,k'}(\omega)=-2 \Im G^{r}_{\alpha,k,k'}(\omega)$, we get:

\begin{align}
J_{\alpha} = \frac{e \Gamma_{\alpha}}{2} \, \left( 1 + \int \frac{d \omega}{2 \pi} \text{\bf Tr} \left[ \left( \underline{\sigma}^N - \underline{\sigma}^1\right) \circ \Im \underline{G}^{<}_{\alpha} (\omega)\right]  \right).
\label{jj_0_causal}
\end{align}

One can show that the transport properties of the chain in the absence of light-matter coupling are only driven by the ratio between $\Gamma_{\alpha}$ and $t_{\alpha}$, i.e. that the steady-state current does not depend on the chain length $N$. Therefore, we now set $N=2$ and $g=0$ in the equations of Secs.~\ref{ss_cucu_sec}, \ref{ee_GF_funsec}, and \ref{ph_sec_app}, and after some straightforward algebra, we can write Eq.~(\ref{jj_0_causal}) as:

\begin{align}
J_{\alpha} &= \frac{e \Gamma_{\alpha}}{2} \, \left( 1 - \int \frac{d \omega}{2 \pi} \frac{\Gamma \left( (\omega + t_{\alpha}) (\omega - t_{\alpha}) + \Gamma_{\alpha}^2/4 \right)}{ \left((\omega + t_{\alpha})^2 + \Gamma_{\alpha}^2/4 \right) \left((\omega - t_{\alpha})^2 + \Gamma_{\alpha}^2/4\right)} \right) \nonumber \\
&= \frac{e \Gamma_{\alpha}}{2} \, \left( 1 - \frac{\Gamma_{\alpha}^2 t_{\alpha}/4}{\frac{\Gamma_{\alpha}^2 t_{\alpha}}{4} + t_{\alpha}^3} \right) \nonumber \\
&= \frac{e \Gamma_{\alpha}/2}{1 + \left(\frac{\Gamma_{\alpha}}{2t_{\alpha}}\right)^2}.
\label{jj_1_causal}
\end{align}

When $t_{\alpha} \ll \Gamma_{\alpha}$, many electrons travel through the chain at the same time, the transport is thus inhibited due to the Pauli principle, and $J_{\alpha} \sim 2 e t^{2}_{\alpha}/\Gamma_{\alpha} \ll e\Gamma_{\alpha}/2$. In the opposite regime $t_{\alpha} \gg \Gamma_{\alpha}$, electrons travel one by one, and the current then reaches its maximum $J_{\alpha} \sim e\Gamma_{\alpha}/2$. In this case, the electron DOS (as well as the transmission $T_{\alpha}(\omega)$) exhibits $N$ well-resolved peaks of width $\sim \Gamma_{\alpha}$ distributed over the bandwidth $4 t_{\alpha}$, corresponding to the chain Bloch states.  

\subsection{Second order spectral broadening}
\label{sd_or_spec}

The scaling of the electron spectral broadening induced by the light-matter coupling can be found by evaluating analytically the second-order ($\propto g^{2}$) electron SE. As explained in the beginning of Sec.~\ref{self_cons_proc}, the latter is obtained by replacing $\underline{G}^{\lessgtr}_{\alpha}$ by $\underline{G}^{0\lessgtr}_{\alpha}$, and $D^{\lessgtr}$ by $D^{\lessgtr}_{1}$ in Eq.~(\ref{lm_se_lala}). At resonance $\omega_{0}=\omega_{21}=1$, it becomes:

\begin{align} 
\chi_{\alpha,k,k'} (\omega)= \frac{4\kappa g^{2} (1-\delta_{\alpha,\alpha'})\delta_{k,k'}}{\left((\omega-\omega_{\alpha',k})^{2}-1\right)^{2}+\kappa^{2}} \Big(N_{\rm ph}+ (1- n^{0}_{\alpha',k} ) \theta (\omega-\omega_{\alpha',k}) + n^{0}_{\alpha',k}  \theta (\omega_{\alpha',k}-\omega)\Big) + \Gamma_{\alpha} \left(\sigma^{N}_{k,k'} - \sigma^{1}_{k,k'}\right),
\label{broad_causal} 
\end{align}

where $n^{0}_{\alpha,k}$ denotes the Fermi occupation number of the Bloch state $k$ in the band $\alpha'$, for $g=0$. Letting the leads contribution aside [second term in the right-hand side of Eq.~(\ref{broad_causal})], the broadening is diagonal with respect to $k$. Considering a Bloch state $k$ in the lowest band $\alpha=1$, its light-induced broadening depends on the filling of the state $k$ in the upper band $\alpha'=2$. When $n_{2,k}=1$, the associated electron can undergo a transition from the upper to the lower band by emitting a photon with energy $\omega_{0}=\omega_{21}$ (at resonance). In the vicinity of $\omega_{1}$, one has $\omega_{2,k} - \omega \approx 1$, which yields: 

\begin{align} 
\chi_{1,k} \sim \frac{4 g^{2}}{\kappa} \left(N_{\rm ph} + n^{0}_{2,k} \right).
\label{broad_causal2} 
\end{align}

When the state $(2,k)$ in the upper band is empty $n^{0}_{2,k}=0$, the light-induced broadening of Bloch states in the lower band is thus only possible in the presence of a finite photon population. This explains why when $\Gamma_{2}$ becomes very small (poor injection/exctraction in the upper band), $n_{2,k}$ gets close to zero, and one crucially needs $N_{\rm ph}\neq 0$ to obtain an efficient broadening $\sim g^{2}/\kappa$ which provides the full restoration of the current flowing in the lower band. In this case, also with $t_{2} \gg \Gamma_{2}$, the maximum current enhancement is given by:

\begin{align}
\frac{J}{J_{0}} \sim \frac{\left[ 1  + \frac{\Gamma_2}{\Gamma_1}\right]}{\left[ \left(\frac{2 t_1}{\Gamma_1} \right)^2  + \frac{\Gamma_2}{\Gamma_1}\right]} \approx \frac{\Gamma_{1}}{\Gamma_{2}}, \qquad \textrm{for} \quad \Gamma_{2} \ll \Gamma_{1} \quad \textrm{and} \quad t_{1} \ll \Gamma_{1}.
\label{eq:stb2}
\end{align}

\subsection{Collective effects} 
\label{bobo_enh_sec}

\begin{figure}[ht]
\centerline{\includegraphics[width=300pt]{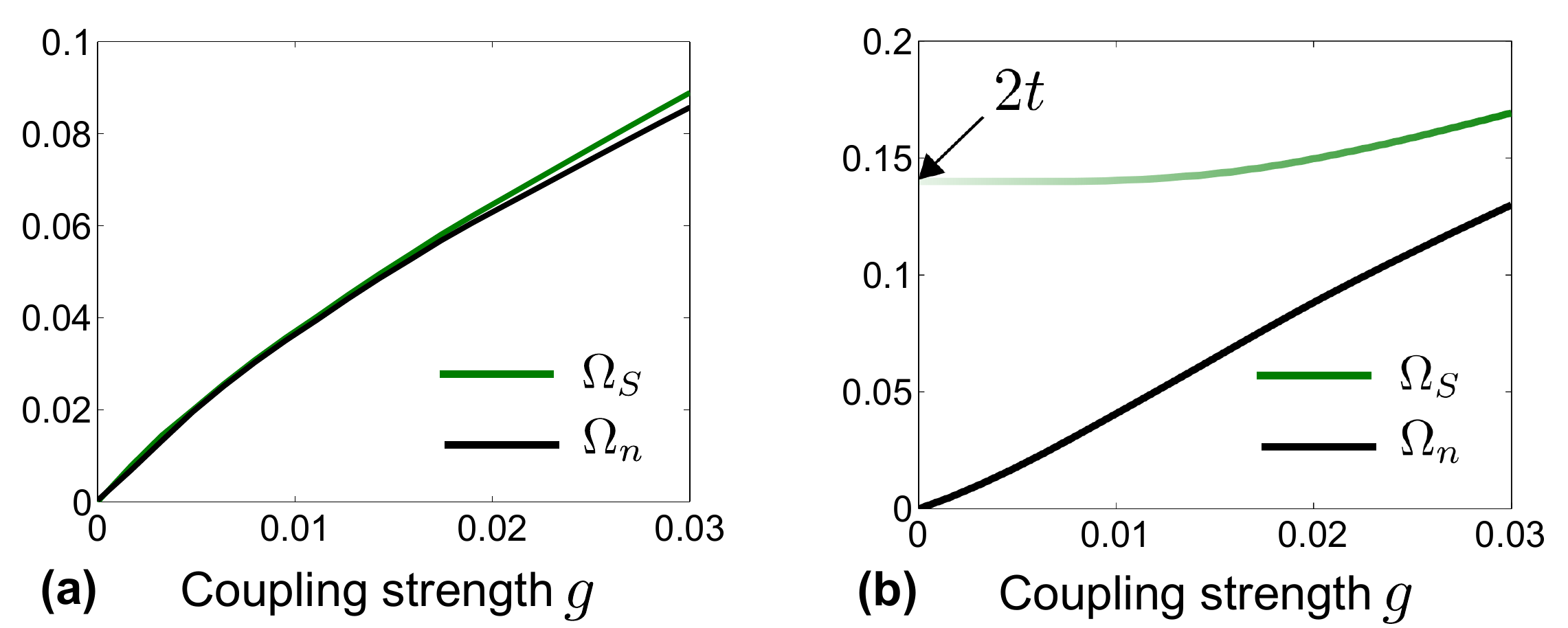}}
\caption{$\Omega_S$ and $\Omega_n$ versus $g$ for $N=30$ and $\Gamma=2.5 \times 10^{-4}$. \textbf{(a)} Collective dressing regime with $t=2.5 \times 10^{-3}$ and $\kappa=0.05$. \textbf{(b)} Individual dressing regime with $t=0.07$ and $\kappa=5 \times 10^{-3}$.}
\label{fig4}
\end{figure}

To further display the crossover between individual and collective regimes discussed in the main text, we compare $\Omega_S$ to $\Omega_{n}$, and summarize the results in Fig.~\hyperref[fig4]{\ref*{fig4}}, in the case $t_{1}=0$, $t_{2}\equiv t$, and $\Gamma_{1}=\Gamma_{2}\equiv \Gamma$. In the collective dressing regime [panel\textbf{(a)}], $\Omega_{n}\approx \Omega_S$, showing that the steady-state dynamics partly consists of collective oscillations of the charge density between the two bands with frequency $\Omega_n$. As explained in the main text, this mode does not contribute to the charge transport and competes with the individual dynamics, which results in reducing the current enhancement. In the individual dressing regime [panel\textbf{(b)}], $\Omega_{S}$ and $\Omega_{n}$ clearly differ, indicating that the individual dressing of Bloch states associated to the effective hopping mecanism with typical rate $\propto g^{2}/\kappa$ (responsible for the current enhancement) largely dominates. However, when $g \gtrsim 4t$ [not shown in Fig.~\hyperref[fig4]{\ref*{fig4}}\textbf{(b)}], $\Omega_{S}$ and $\Omega_{n}$ coincide which results in recovering the physics of the collective dressing regime.

\subsection{Scaling of the current enhancement} 
\label{sca_sec}

\begin{figure}[ht]
\centerline{\includegraphics[width=330pt]{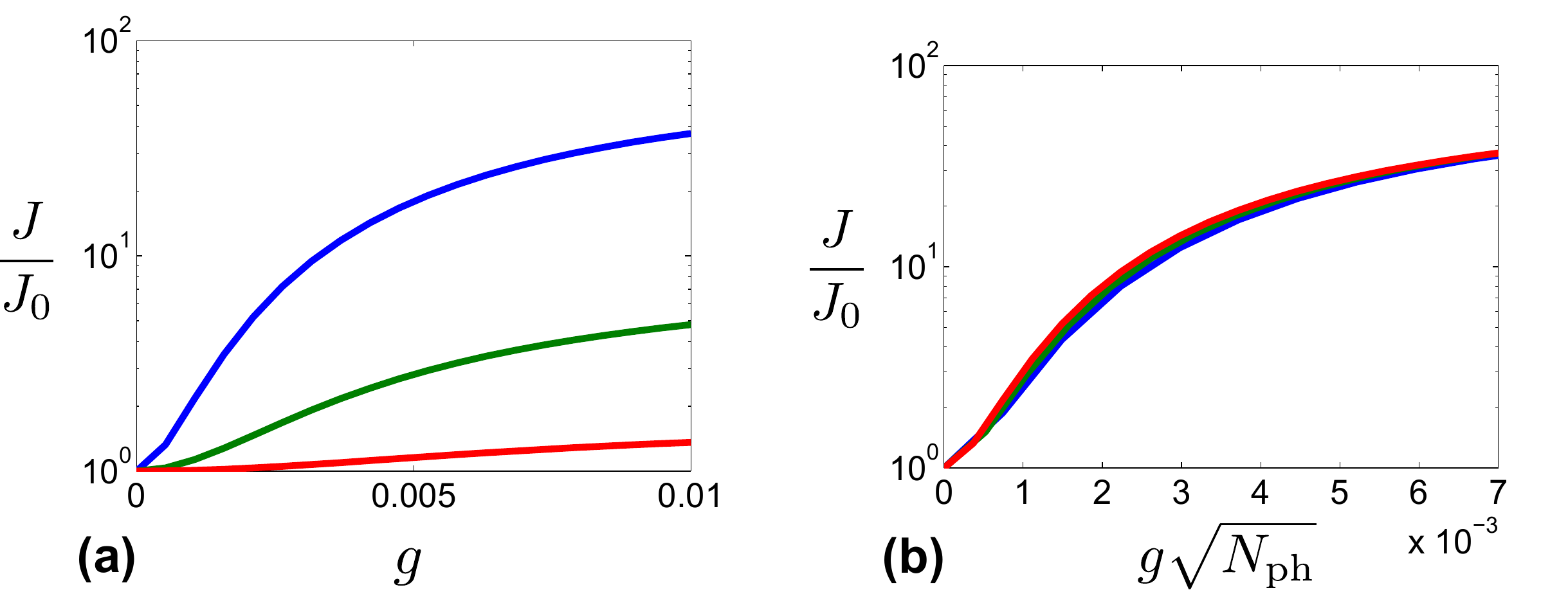}}
\caption{\textbf{(a)}: Current enhancement $J/J_{0}$ versus $g$ for $\Gamma_{2}=10^{-2}$ (red), $\Gamma_{2}=10^{-3}$ (green), $\Gamma_{2}=10^{-4}$ (blue), and $N_{\rm ph}=0.5$. \textbf{(b)}: Current enhancement $J/J_{0}$ versus $g \sqrt{N_{\rm ph}}$ for $N_{\rm ph}=0.5$ (red), $N_{\rm ph}=1$ (green), $N_{\rm ph}=2$ (blue), and $\Gamma_{2}=10^{-4}$. Other parameters are $N=3$, $\Gamma_{1}=10^{-2}$, $t_{1}=5\times10^{-5}$, $t_{2}=7\times 10^{-2}$, $\kappa=5\times 10^{-3}$.}
\label{scale_bos}
\end{figure}

When $\Gamma_{1} \neq \Gamma_{2}$, it is interesting to see how the current enhancement $J/J_{0}$ scales with $\Gamma_{2}$, as well as with the population in the photon bath $N_{\rm ph}$. This is represented on Fig.~\ref{scale_bos}. Panel \textbf{(a)} shows $J/J_{0}$ as a function of the coupling strength $g$ for $N_{\rm ph}=0.5$, $\Gamma_{2}=10^{-2}$ (red), $\Gamma_{2}=10^{-3}$ (green), and $\Gamma_{2}=10^{-4}$ (blue). This confirms the rough estimate of the maximum current enhancement $\Gamma_{1}/\Gamma_{2}$, given at the end of Sec.~\ref{sd_or_spec}. Panel \textbf{(b)} displays $J/J_{0}$ as a function of the collective ``photon-enhanced'' coupling strength $g \sqrt{N_{\rm ph}}$ for $N_{\rm ph}=0.5$ (red), $N_{\rm ph}=1$ (green), and $N_{\rm ph}=2$ (blue). The coincidence of the results for different photon populations shows that the current enhancement for $N_{\rm ph}\neq 0$ corresponds to a bosonic stimulation, as confirmed by the second-order estimation in Eq.~(\ref{broad_causal2}). 

\subsection{Comparison with a quantum master equation model}
\label{comp_app_meth}

It is interesting to compare the results obtained using NEGFs with a method based on the exact numerical solving of the quantum master equation (up to a certain photon number considered as a cutoff in the system's Hilbert space). In this case, the effect of the (Markovian) environment is cast in terms of dissipators, providing both the injection/extraction of cavity photons (scaling with the rate $\kappa$), and electrons in the chain orbitals $\alpha$, with rate $\Gamma_{\alpha}$. The corresponding quantum master equation reads $\partial_t \hat{\rho} = - i [H_S, \hat{\rho}] + \mathcal{L} \hat{\rho}$, where

\begin{align}
\mathcal{L} \hat{\rho} = \frac{\kappa}{2} \left(1+ N_{\rm ph}\right)\mathcal{D} [a] \hat{\rho} + \frac{\kappa}{2} N_{\rm ph} \mathcal{D} [a^{\dagger}] \hat{\rho}+  \sum_{\alpha=1}^{2} \frac{\Gamma_{\alpha}}{2} \left( \mathcal{D}[c^{\dagger}_{\alpha,1}] \hat{\rho} + \mathcal{D} [c_{\alpha,N}] \hat{\rho} \right)
\end{align}

determines the time evolution of the system density operator $\hat{\rho}$, with the Lindblad terms:

\begin{align}
\mathcal{D}[b] \hat{\rho} = - \{b^{\dagger} b, \hat{\rho}\} + 2 b \hat{\rho} b^{\dagger}.
\end{align}

In the steady state, $\partial_t \hat{\rho}=0$, and the current can be written as:

\begin{align}
J = \sum_{\alpha}\textrm{Tr}
\left[\frac{e\Gamma_{\alpha}}{2} c^{\dagger}_{\alpha,1} c_{\alpha,1} \{\mathcal{D} [c^{\dagger}_{\alpha,{1}}] \hat{\rho} \} \right],
\end{align}

where $\textrm{Tr}$ denotes the trace on the system Hilbert space. Due to the exponential scaling of the Hilbert space size with the chain length, this method can be only applied for small systems ($N \sim 5$). Nevertheless, an effective master equation model where the photonic degrees of freedom are adiabatically eliminated can be used for $N > 5$, when the cavity photon decay rate is large compared to the other energy scales. Further, a rate equation model directly derived from the quantum master equation might be promising to tackle larger systems ($N \sim 100$). 

\begin{figure}[t]
\centerline{\includegraphics[width=300pt]{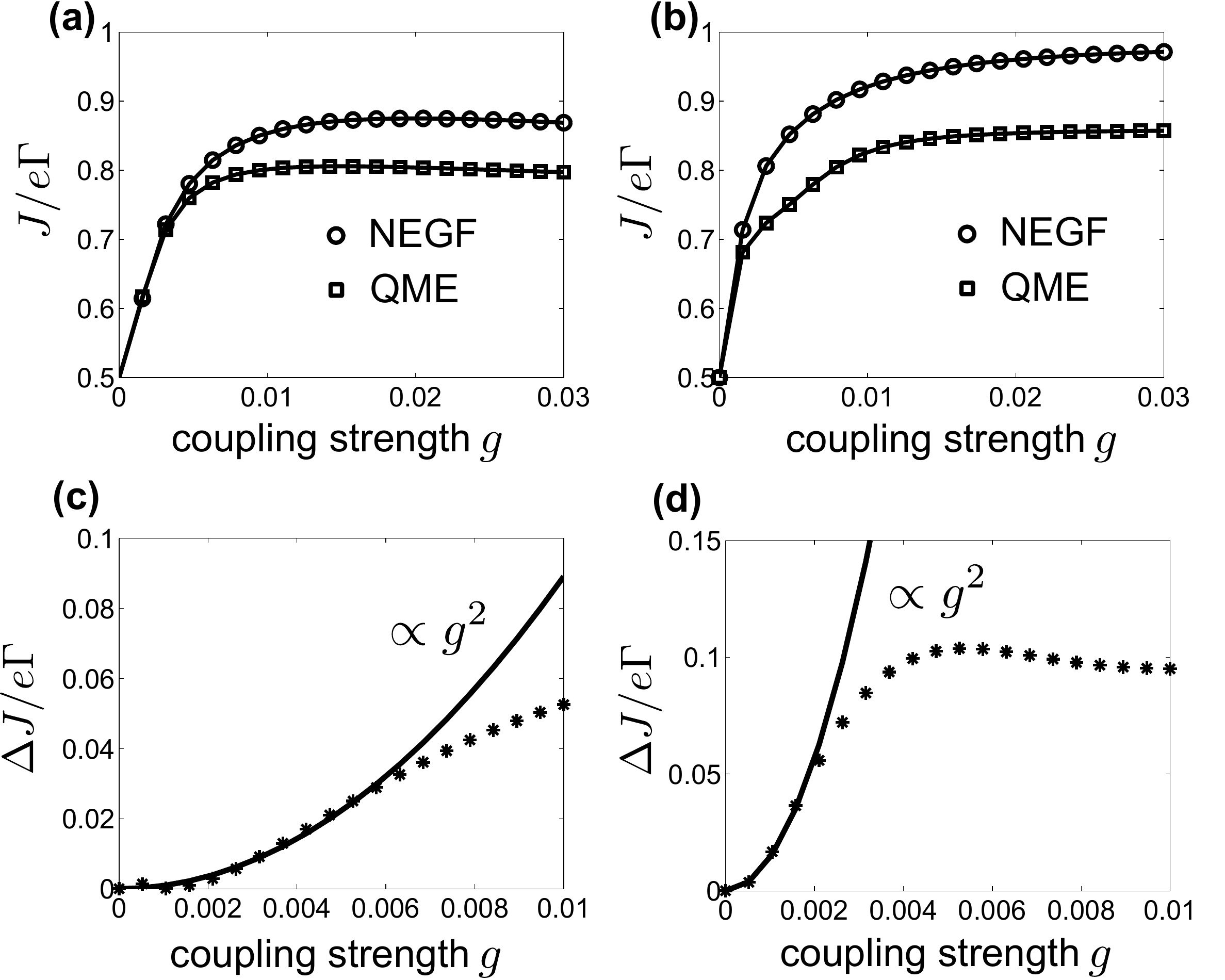}}
\caption{Comparison between the results obtained with the NEGF and the quantum master equation approach (QME). Top panels: Steady-state current $J/(e\Gamma)$ as a function of $g$ in \textbf{(a)} the collective dressing regime with $N=3$, $\Gamma=2.5 \times 10^{-4}$, $t=2.5 \times 10^{-3}$, $\kappa=0.05$, \textbf{(b)} the individual dressing regime with $N=3$, $\Gamma=2.5 \times 10^{-4}$, $t=0.07$, $\kappa=0.005$. Bottom panels: Absolute difference $\Delta J/(e\Gamma)$ between the two methods as a function of $g$, in \textbf{(c)} the collective dressing regime, \textbf{(d)} the individual dressing regime.}
\label{comp}
\end{figure}

On Fig. \ref{comp}, we have represented a comparison between the results obtained with the NEGF and the quantum master equation approaches (QME), in the situation where $t_{1}=0$, $t_{2}\equiv t$, and $\Gamma_{1}=\Gamma_{2}\equiv \Gamma$. The steady-state current $J/(e\Gamma)$ is represented as a function of $g$ in the collective dressing regime with $N=3$, $\Gamma=2.5 \times 10^{-4}$, $t=2.5 \times 10^{-3}$, $\kappa=0.05$ [panel \textbf{(a)}], and in the individual dressing regime with $N=3$, $\Gamma=2.5 \times 10^{-4}$, $t=0.07$, $\kappa=0.005$ [panel \textbf{(b)}]. The absolute difference $\Delta J/(e\Gamma)$ between the two methods is represented in the bottom panels, in \textbf{(c)} the collective dressing regime and \textbf{(d)} the individual dressing regime. The validity of the NEGF model depends on the smallness of the perturbative parameter $g^{2}/(\Gamma \kappa)$. As expected, the absolute difference between the results obtained by the two methods scales as $g^{2}$ for small $g$. By evaluating the perturbative parameter $g^{2}/(\Gamma \kappa)$ when $g$ is on the high side in the considered range, one realizes that the former is rather large for $g\sim 0.03$, which explains the discrepancy ($\Delta J\sim 0.1 e\Gamma$) between the two methods in this range. However, the qualitative trends are unchanged, and the NEGF model is still expected to give a correct physical description of the system, even beyond the boundaries of this perturbative approach. 

\begin{figure}[t]
\centerline{\includegraphics[width=300pt]{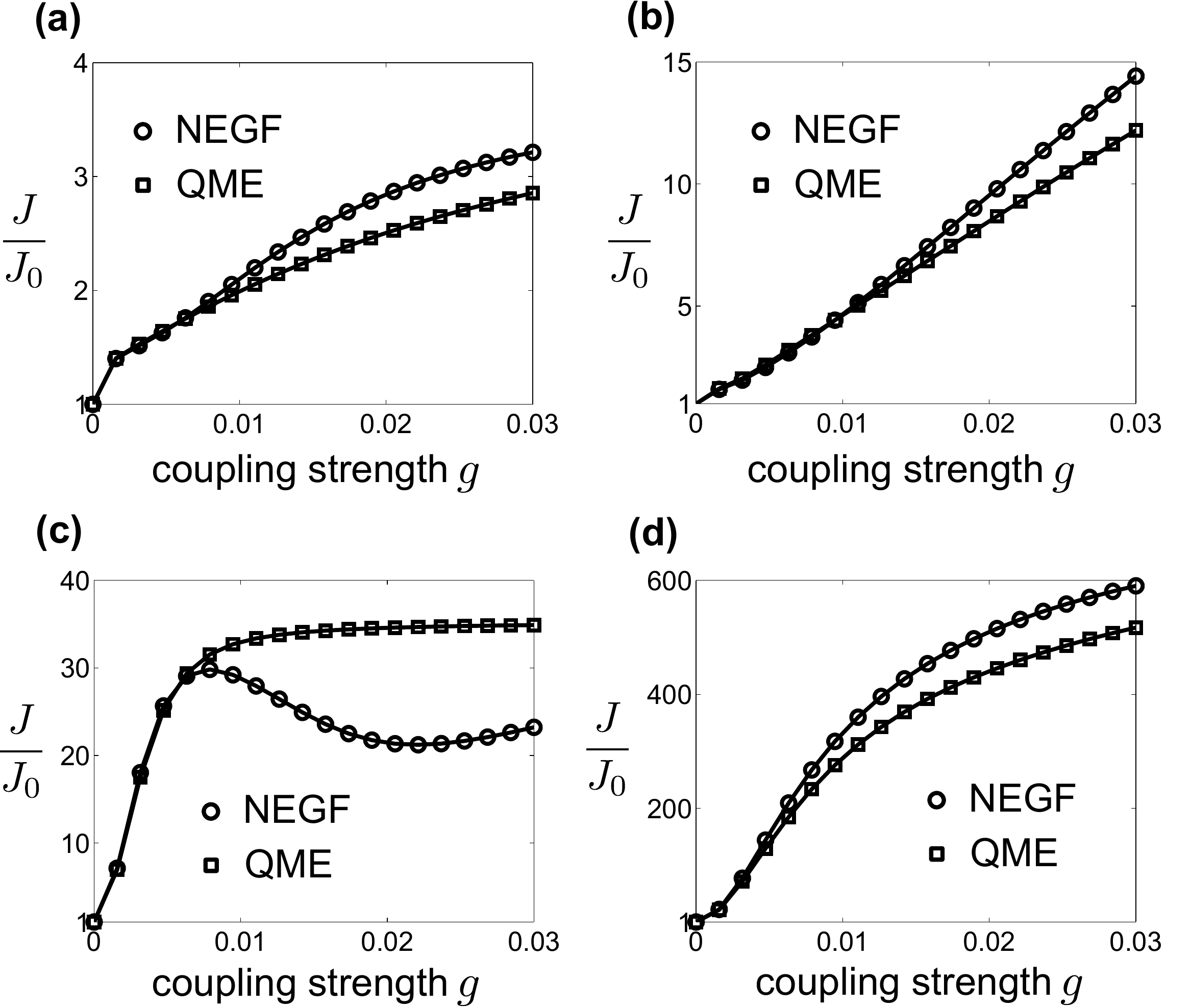}}
\caption{Comparison between the current enhancement $J/J_{0}$ as a function of $g$ obtained with the NEGF and the quantum master equation approach (QME), for $\Gamma_{1}\neq \Gamma_{2}$. Top panels: $N_{\rm ph}=0$, \textbf{(a)} collective dressing regime with $t_{2}=2.5 \times 10^{-3}$, $\kappa=0.05$, and $\Gamma_{1}=10^{-3}$, and \textbf{(b)} individual dressing regime with $t_{2}=0.07$, $\kappa=0.005$, and $\Gamma_{1}=10^{-2}$. Bottom panels: $N_{\rm ph}=0.5$, \textbf{(c)} collective dressing regime with $t_{2}=2.5 \times 10^{-3}$, $\kappa=0.05$, and $\Gamma_{1}=10^{-3}$, and \textbf{(d)} individual dressing regime with $t_{2}=0.07$, $\kappa=0.005$, and $\Gamma_{1}=10^{-2}$. Other parameters are: $N=3$, $\Gamma_{2}=10^{-5}$, $t_{1}=5 \times 10^{-5}$.}
\label{comp2}
\end{figure}

On Fig. \ref{comp2}, we compare the two methods in the regime where $\Gamma_{1}\neq \Gamma_{2}$, and $N_{\rm ph} \geq 0$, both in the collective and individual dressing regimes. Except for panel \textbf{(b)}, showing a spurious hump for $g \approx 0.01$ with the NEGF method, the results obtained with the two methods are in good qualitative agreement.

\end{document}